\documentclass[twocolumn]{aastex63} 

\usepackage{CJKutf8}
\usepackage{pbox}
\usepackage{footnotebackref}
\usepackage{subfigure}
\usepackage{makecell}
\usepackage{amsmath}
\usepackage{multirow}
\usepackage{natbib}
\shorttitle{Evolution of cool gas in CGM and IGM}
\shortauthors{Xuanyi Wu et al.}

\begin{document}

\makeatletter
\LetLtxMacro{\BHFN@Old@footnotemark}{\@footnotemark}

\renewcommand*{\@footnotemark}{%
    \refstepcounter{BackrefHyperFootnoteCounter}%
    \xdef\BackrefFootnoteTag{bhfn:\theBackrefHyperFootnoteCounter}%
    \label{\BackrefFootnoteTag}%
    \BHFN@Old@footnotemark
}
\makeatother

\title{Tracing the evolution of the cool gas in CGM and IGM environments through Mg II absorption from redshift z=0.75 to z=1.65 using DESI-Y1 data}

\correspondingauthor{Xuanyi Wu}
\email{xuanyiwu1996@gmail.com}
\author{Xuanyi Wu}
\affiliation{Department of Astronomy, Tsinghua University, Beijing 100084, China}

\author[0000-0001-8467-6478]{Z.~Cai}
\affiliation{Department of Astronomy, Tsinghua University, Beijing 100084, China}

\author[0000-0001-8857-7020]{T.-W.~Lan}
\affiliation{Graduate Institute of Astrophysics and Department of Physics, National Taiwan University, No. 1, Sec. 4, Roosevelt Rd., Taipei 10617, Taiwan}
\author[0000-0002-3983-6484]{S.~Zou}
\affiliation{Department of Astronomy, Tsinghua University, Beijing 100084, China}
\affiliation{Chinese Academy of Sciences South America Center for Astronomy, National Astronomical Observatories, CAS, Beijing 100101, China}
\author{A.~Anand}
\affiliation{Lawrence Berkeley National Laboratory, 1 Cyclotron Road, Berkeley, CA 94720, USA}
\author[0000-0002-5665-7912]{Biprateep~Dey}
\affiliation{Department of Physics \& Astronomy and Pittsburgh Particle Physics, Astrophysics, and Cosmology Center (PITT PACC), University of Pittsburgh, 3941 O'Hara Street, Pittsburgh, PA 15260, USA}
\author{Z.~Li}
\affiliation{Department of Astronomy, Tsinghua University, Beijing 100084, China}

\author{J.~Aguilar}
\affiliation{Lawrence Berkeley National Laboratory, 1 Cyclotron Road, Berkeley, CA 94720, USA}

\author[0000-0001-6098-7247]{S.~Ahlen}
\affiliation{Physics Dept., Boston University, 590 Commonwealth Avenue, Boston, MA 02215, USA}

\author{D.~Brooks}
\affiliation{Department of Physics \& Astronomy, University College London, Gower Street, London, WC1E 6BT, UK}

\author{T.~Claybaugh}
\affiliation{Lawrence Berkeley National Laboratory, 1 Cyclotron Road, Berkeley, CA 94720, USA}

\author[0000-0002-1769-1640]{A.~de la Macorra}
\affiliation{Instituto de F\'{\i}sica, Universidad Nacional Aut\'{o}noma de M\'{e}xico,  Cd. de M\'{e}xico  C.P. 04510,  M\'{e}xico}

\author{P.~Doel}
\affiliation{Department of Physics \& Astronomy, University College London, Gower Street, London, WC1E 6BT, UK}

\author[0000-0003-4992-7854]{S.~Ferraro}
\affiliation{Lawrence Berkeley National Laboratory, 1 Cyclotron Road, Berkeley, CA 94720, USA}
\affiliation{University of California, Berkeley, 110 Sproul Hall \#5800 Berkeley, CA 94720, USA}

\author[0000-0002-2890-3725]{J.~E.~Forero-Romero}
\affiliation{Departamento de F\'isica, Universidad de los Andes, Cra. 1 No. 18A-10, Edificio Ip, CP 111711, Bogot\'a, Colombia}
\affiliation{Observatorio Astron\'omico, Universidad de los Andes, Cra. 1 No. 18A-10, Edificio H, CP 111711 Bogot\'a, Colombia}

\author[0000-0003-3142-233X]{S.~Gontcho A Gontcho}
\affiliation{Lawrence Berkeley National Laboratory, 1 Cyclotron Road, Berkeley, CA 94720, USA}

\author{K.~Honscheid}
\affiliation{Center for Cosmology and AstroParticle Physics, The Ohio State University, 191 West Woodruff Avenue, Columbus, OH 43210, USA}
\affiliation{Department of Physics, The Ohio State University, 191 West Woodruff Avenue, Columbus, OH 43210, USA}
\affiliation{Department of Physics, The Ohio State University, 191 West Woodruff Avenue, Columbus, OH 43210, USA}
\affiliation{The Ohio State University, Columbus, 43210 OH, USA}

\author{S.~Juneau}
\affiliation{NSF NOIRLab, 950 N. Cherry Ave., Tucson, AZ 85719, USA}

\author{R.~Kehoe}
\affiliation{Department of Physics, Southern Methodist University, 3215 Daniel Avenue, Dallas, TX 75275, USA}

\author[0000-0003-3510-7134]{T.~Kisner}
\affiliation{Lawrence Berkeley National Laboratory, 1 Cyclotron Road, Berkeley, CA 94720, USA}

\author{A.~Lambert}
\affiliation{Lawrence Berkeley National Laboratory, 1 Cyclotron Road, Berkeley, CA 94720, USA}

\author[0000-0003-1838-8528]{M.~Landriau}
\affiliation{Lawrence Berkeley National Laboratory, 1 Cyclotron Road, Berkeley, CA 94720, USA}

\author[0000-0001-7178-8868]{L.~Le~Guillou}
\affiliation{Sorbonne Universit\'{e}, CNRS/IN2P3, Laboratoire de Physique Nucl\'{e}aire et de Hautes Energies (LPNHE), FR-75005 Paris, France}

\author[0000-0003-4962-8934]{M.~Manera}
\affiliation{Departament de F\'{i}sica, Serra H\'{u}nter, Universitat Aut\`{o}noma de Barcelona, 08193 Bellaterra (Barcelona), Spain}
\affiliation{Institut de F\'{i}sica d’Altes Energies (IFAE), The Barcelona Institute of Science and Technology, Campus UAB, 08193 Bellaterra Barcelona, Spain}

\author[0000-0002-1125-7384]{A.~Meisner}
\affiliation{NSF NOIRLab, 950 N. Cherry Ave., Tucson, AZ 85719, USA}

\author{R.~Miquel}
\affiliation{Instituci\'{o} Catalana de Recerca i Estudis Avan\c{c}ats, Passeig de Llu\'{\i}s Companys, 23, 08010 Barcelona, Spain}
\affiliation{Institut de F\'{i}sica d’Altes Energies (IFAE), The Barcelona Institute of Science and Technology, Campus UAB, 08193 Bellaterra Barcelona, Spain}

\author[0000-0002-2733-4559]{J.~Moustakas}
\affiliation{Department of Physics and Astronomy, Siena College, 515 Loudon Road, Loudonville, NY 12211, USA}

\author{J.~ A.~Newman}
\affiliation{Department of Physics \& Astronomy and Pittsburgh Particle Physics, Astrophysics, and Cosmology Center (PITT PACC), University of Pittsburgh, 3941 O'Hara Street, Pittsburgh, PA 15260, USA}

\author[0000-0001-7145-8674]{F.~Prada}
\affiliation{Instituto de Astrof\'{i}sica de Andaluc\'{i}a (CSIC), Glorieta de la Astronom\'{i}a, s/n, E-18008 Granada, Spain}

\author{G.~Rossi}
\affiliation{Department of Physics and Astronomy, Sejong University, Seoul, 143-747, Korea}

\author[0000-0002-9646-8198]{E.~Sanchez}
\affiliation{CIEMAT, Avenida Complutense 40, E-28040 Madrid, Spain}

\author{D.~Schlegel}
\affiliation{Lawrence Berkeley National Laboratory, 1 Cyclotron Road, Berkeley, CA 94720, USA}
\author{M.~Schubnell}
\affiliation{Department of Physics, University of Michigan, Ann Arbor, MI 48109, USA}
\affiliation{University of Michigan, Ann Arbor, MI 48109, USA}

\author[0000-0002-2949-2155]{M.~Siudek}
\affiliation{Institute of Space Sciences, ICE-CSIC, Campus UAB, Carrer de Can Magrans s/n, 08913 Bellaterra, Barcelona, Spain}

\author{D.~Sprayberry}
\affiliation{NSF NOIRLab, 950 N. Cherry Ave., Tucson, AZ 85719, USA}
\author[0000-0003-1704-0781]{G.~Tarl\'{e}}
\affiliation{University of Michigan, Ann Arbor, MI 48109, USA}
\author{B.~A.~Weaver}
\affiliation{NSF NOIRLab, 950 N. Cherry Ave., Tucson, AZ 85719, USA}
\author[0000-0002-6684-3997]{H.~Zou}
\affiliation{National Astronomical Observatories, Chinese Academy of Sciences, A20 Datun Rd., Chaoyang District, Beijing, 100012, P.R. China}

\begin{abstract}

We present a measurement of the mean absorption of cool gas traced by Mg II (${\lambda\lambda 2796, 2803}$) around emission line galaxies (ELGs), spanning spatial scales from 20 kpc to 10 Mpc. The measurement is based on cross-matching the positions of about 2.5 million 
ELGs at $z = 0.75-1.65$ and the metal absorption in the spectra of 1.4 million background quasars with data provided by the Year 1 sample of the Dark Energy Spectroscopic Instrument (DESI). The ELGs are divided into two redshift intervals: $0.75 < z < 1.0$ and $1.0 < z < 1.65$. We find that the composite spectra constructed by stacking the ELG-QSO pairs show evolution with redshift, with $z>1$ having a systematically higher signal of Mg II absorption. Within 1 Mpc, the covering fraction of 
the cool gas at $z > 1$ is higher than that of $z < 1$. The enhancement becomes less apparent especially if the projected distance $r_{p}>$1 Mpc. 
Also, ELGs with higher stellar mass and star formation rate (SFR) yield higher clustering of Mg II absorbers at $z<1$. For $z>1$, the covering fractions with different SFRs show little difference. The higher Mg II absorption at higher redshift also supports the observations of higher star formation at cosmic noon. Besides,  the profile of Mg II absorption reveals a change of slope on scales of about 1 Mpc, consistent with the expected transition from a dark matter halo-dominated environment to a regime where clustering is dominated by halo-halo correlations. We estimate the cool gas density profile and derive the metal abundance at different redshifts. The growth of metal abundance suggests an increased presence of cool gas in the intergalactic medium (IGM) towards higher redshifts. 

\end{abstract}

\section{Introduction}
The $\Lambda$CDM model suggests that galaxies form within dark matter halos through gas cooling and condensation. The circumgalactic medium (CGM) represents the gas environment surrounding galaxies, where gas accretion and outflow occur. It is a vital component for understanding galaxy formation and evolution (\citealt{tumlinson2017}, for a review). The intergalactic medium (IGM) reflects the larger-scale distribution of gas and connects with galaxies through the CGM. Gas is accreted from the IGM into galaxies, while some of it is expelled from galaxies due to stellar and active galactic nucleus (AGN) feedback processes \citep{Katz2003,keres2005,Dekel2006}. The precise mechanisms by which the gas regulates star formation and quenching of galaxies are still not fully understood. And the cycling of multiphase gas makes the baryon exchanges in the CGM and IGM a complex process \citep{Martizzi2019,peroux2020}.

Investigations of various galaxy-absorber systems have revealed the distribution and dynamics of gas absorbers around galaxies. Absorbers are identified by searching in the spectra of background sources, such as quasars \citep[]{Kacprzak2008,menard2009,menard2011,nielsen2013,anand2021,Napolitano_2023}. Cross-correlating absorbers with foreground galaxies provides information on the gas environment around galaxies at scales of hundreds of kiloparsecs \citep[e.g.][]{lovegrove2011,lundren2011,Rubin2018}. Previous studies have shown that the rest-frame equivalent width of Mg \textsc{ii} and the impact parameter of absorbers and galaxies are anti-correlated and give the mean covering fraction of Mg II absorbers a standard halo gas radius spanning different stellar mass ranges \citep[e.g.][]{Kacprzak2008, chen2010b,bordoloi2011,chenhw2010}. Within a distance of 100 kpc, the covering fraction of Mg II absorbers with rest-frame equivalent width larger than 0.3$\rm{\AA}$ around galaxies is approximately 80$\%$, which decreases significantly when extending to larger distances \citep[e.g.][]{chenhw2008}. Using VLT/MUSE, many groups have selected samples of Mg II absorbers around galaxies at $z\approx1$ and traced the gas flows as well as the gas abundance in halos that serves as environment \citep{Zabl2019,Schroetter2019,Rajeshwari2020,zousw2024}. Stacking the spectra at the rest frame of foreground galaxies helps to increase the signal-to-noise ratio (SNR) of metal absorption. Additionally, Wu et al. in prep, hereafter Wu24, have shown that taking the average signal of each galaxy-quasar pair directly provides equivalent results. Both statistical analysis of absorbers and stacked absorption profiles offer valuable information related to the scaling relations between gas and galaxies \citep[]{pieri2010,zhu2013,zhu2014,lan2014,pieri2014,ignasi2015,murga2015,huang2016}.

Decades of research have contributed to an improved understanding of the gas environment and galaxy evolution, yet the intrinsic role of CGM/IGM remains unclear. Therefore, it is necessary to investigate the gas properties in the CGM and IGM surrounding galaxies with different characteristics, such as varying SFR and stellar masses. Studies on various galaxy-absorber systems have provided insights into the distribution and dynamics of gas absorbers around galaxies. Galaxies with higher SFR tend to have a higher abundance of strong absorbers. Also, the distribution of absorbers in the CGM exhibits anisotropy, which aligns with current galaxy formation models \citep{bordoloi2011,menard2011,lan2020,Dutta2021,Zou_2024}. Using cross-correlation between quasar absorbers and foreground galaxies, previous studies have shown that passive galaxies have lower mean absorption and covering fraction of cool gas traced by Mg II absorbers compared with star-forming galaxies which are less massive \citep{Lan2018,anand2021,Huang2021,Anand2022}. Simulations conducted from local to high redshifts have revealed the co-evolution of halos and galaxies, positioning cool gas of star-forming galaxies as valuable sources of information regarding star formation feedback and environmental effects \citep{diemand2008,vogelsberger2020}. 

The Dark Energy Spectroscopic Instrument \citep[DESI; ][]{Snowmass2013Levi,DESI2016A,DESI2016B,DESI2022} is a new large survey designed to provide a comprehensive map of the universe, including the distribution of gas in and around galaxies. DESI will yield a large sample of quasar spectra and emission-line galaxies (ELGs) at higher redshifts than previous surveys such as SDSS, enabling the detection of gas distribution over greater distances and with higher precision. In this case, DESI would provide a comprehensive understanding of galaxy evolution and the role of gas in their formation and evolution at higher redshift \citep{desiwhite2016}.

We describe our data in Section \ref{data}. In Section \ref{method}, we introduce our method of normalizing spectra and extract the average absorption signal of MgII. Section \ref{abs_diffz} contains the results of Mg II absorption around ELGs. We present the impact of stellar mass and SFR of galaxies in Section \ref{sfr_mass_z} and derive the column density of cool gas in Section \ref{surfacedensity_siii}. In section \ref{summary} we summarize and discuss our results. 

In this paper, we adopt a flat $\Lambda$CDM cosmology with 
matter density $\Omega_{m}=0.3$ and the reduced Hubble 
constant $h=0.7$. $W_{2796}$ infers to the rest-frame equivalent width of Mg II $\lambda$2796 line. $W_0$ infers to the rest-frame equivalent width of Mg II doublet, which is the sum of equivalent widths of Mg II $\lambda$2796 and Mg II $\lambda$2803.


\section{Data}\label{data}

DESI is a stage IV dark-energy survey that aims to understand better the properties of dark energy and dark matter \citep{DESI2016A,DESI2016B,DESI2022KP1,Silber2023,Miller2023}. It builds on the success of several Stage III experiments, including the BOSS/eBOSS spectroscopic redshift surveys (extended Baryonic Oscillation Spectroscopic Survey; \citealt{dawson2013,dawson2016}) and the DES imaging survey (Dark Energy Survey; \citet{DES2016}). We select the ELGs and QSOs from the DESI internal release ``Iron'', to be publicly released in the DESI Data Release 1 (DR1; DESI Collaboration et al. 2024 I, in prep). The data covers the Commissioning, Survey Validation (SV), Main, and ``special'' survey data from December 14th, 2020 - June 13th 2022. The DR1 will be the first major data release of the DESI collaboration that will include about 10 times more spectra than the early data release which was released on June 2023 \citep{DESI_EDR}.

\subsection{The ELG sample}
 
The ELGs are selected based on their optical and near-infrared photometry. They are mostly star-forming galaxies with strong [OII] $\lambda \lambda 3726,3729$ emission. In this work, we select the ELG sample with $\rm{log_{10}}[\rm{OII}]S/N > 0.9-0.2\times\rm{log_{10}}\Delta \chi^{2}$ and \texttt{ZWARN==0}, which ensures the reliability of the estimation of redshift \citep{Raichoor_2023,Lan2023}. The $[\rm{OII}]S/N$ is the SNR of [OII] $\lambda \lambda 3726,3729$ emission. $\Delta \chi^{2}$ is difference between the $\chi^{2}$ of the best and 
2nd best-fit model. The redshift and $\chi^{2}$ are derived by \textit{Redrock}\footnote{https://github.com/desihub/redrock/} algorithm \citep[S. Bailey et al. 2024, in prep;][]{,anand2024}. Besides, we further select the sample with redshifts in the range of $0.75<z<1.65$ that covers about 2.5 million ELGs. The SDSS eBOSS survey established an ELG sample with a central redshift of approximately 0.85. The ELGs have a high density per square degree in the DESI survey, which makes them excellent tracers of large-scale structures and have a central redshift of about 1.05. The redshift is systematically higher than the eBOSS sample, opening the window to probe the gas properties at $z>1$.

We use the stellar mass of ELG that is estimated based on Random Forests \citep{breiman2001} that maps photometric data from three optical bands ($g,r,z$) as well as WISE $W1$ and $W2$ to Stripe 82 Massive Galaxy Catalog (S82-MGC) stellar masses \citep{bundy2015}. For details of the methodology, please refer to \citet{Zhourp2023}. 

We estimate the SFR of ELGs using EAZY \citep{Brammer2008}, incorporating photometric data from three optical bands ($g,r,z$) and $W1$ and $W2$ bands for each galaxy. Given our primary focus on assessing the correlation between SFR and Mg II, our analysis mainly emphasis on the relative SFR values. The distribution of SFR at different redshifts is shown in Figure \ref{fig:z-sfr} in Appendix \ref{appSFRandmass}, color-coded with stellar mass. The average stellar mass of our ELG sample extends from $10^{9.8}M_{\odot}$ to $10^{10.3}M_{\odot}$ as the redshift changes from 0.8 to 1.6. We also observe an increase in the SFR of ELGs with redshift, with the median value rising from approximately 7 to 20 $M_{\odot}yr^{-1}$ as redshift spans from 0.8 to 1.6.

\subsection{The QSO sample}
We use about 1.4 million QSOs with $\Delta \chi^{2}>15$ as background tracers to remove the unqualified estimation of redshift, which left about 98.5$\%$ of the spectra \cite{Chaussidon2023}. The redshift range of QSOs spans from 0.75 to 5 and has half of the sources with $z>1.6$. The spectroscopic reduction pipeline is developed by \citet{Guy2023}. We obtain ELG-QSO pairs at a certain projected distance by comparing the positions of ELGs and QSOs in the sky and select pairs where $z_{qso}-z_{gal}>0.1$ to make sure that the QSO is behind the galaxy. Moreover, we exclude the cases where Mg II drops into the $Ly\alpha$ forest because estimating the continuum in the $Ly\alpha$ forest has large uncertainty. Thus we only take the pairs with $(1+z_{gal})\times2800\rm{\AA} > (1+z_{qso})\times 1216\rm{\AA}$ in our study. The redshift of QSO is derived by \textit{Redrock} algorithm. The visual inspection and validation of QSOs are presented in \cite{Alexander2023}.


\section{Method overview}\label{method}

In this section, we describe our method for measuring the rest-frame equivalent width (EW) of the Mg II absorption line in ELG-QSO pairs. In Wu24 we establish the QSO continuum fitting procedure and use stacking as well as the force detection method to derive the mean equivalent width profile. Here we briefly summarize our method.

First, we normalize the QSO spectra using eigenspectra from \texttt{non-negative matrix factorization} (NMF) decomposition \citep{Lee1999}, as
implemented by \cite{zhu2016}\footnote{https://github.com/guangtunbenzhu/NonnegMFPy}. After normalization with eigenvectors, we apply median filtering with two filters of widths 71 and 141 pixels iteratively to remove them. The widths of the filters are much larger than the wavelength coverage of Mg II, which is about 20 pixels, thus the filters would remove the fluctuations without smoothing out the Mg II absorption. The iteration stops as the standard deviation of flux within [2750,2795] $\rm{\AA}$ and [2805,2845] $\rm{\AA}$ is smaller than 2 or the iteration time is larger than five. We also perform $\sigma$-clipping with a threshold of 1.5$\sigma$ in earlier iterations and 2.5$\sigma$ in the last iteration to remove the outliers.

Next, we identify ELG-QSO pairs at different projected distances $r_p$, and shift the QSO spectrum to the rest-frame of each ELG. The rest-frame equivalent width of $i$-th galaxy-QSO pair is:
\begin{equation}\label{eq:indi_ew}
W_{0}^{i}\left(r_{p}\right)=\int\left[1-R^{i}\left(\lambda, r_{p}\right)\right] d \lambda 
\end{equation}
where $R^{i}\left(\lambda, r_{p}\right)$ is the continuum normalized spectrum of quasar in the rest-frame of the foreground galaxy, and we integrate the normalized flux from 2789 $\rm{\AA}$ to 2809 $\rm{\AA}$ for MgII $\lambda \lambda 2796,2803$ doublet. 

The average signal of $W_{0}^{i}$ is:

\begin{align}
\left\langle W_{0}\right\rangle\left(r_{p}\right) & =\frac{\sum_{i} w_{i} W_{0}^{i}\left(r_{p}\right)}{\sum_{i} w_{i}}\label{eq:force} \\
& =\frac{1}{\sum_{i} w_{i}} \sum_{i}\left\{w_{i} \cdot \int\left[1-R^{i}\left(\lambda, r_{p}\right)\right] d \lambda\right\}\\
& =\int\left[1-\left\langle R\left(\lambda, r_{p}\right)\right\rangle\right] d \lambda\label{eq:stacking}
\end{align}

The $w_{i}$ is the weight of the $i$-th pair, in this work, we take $w_{i}=1$. Also, we assume that $w_{i}$ remains independent of wavelength for each spectrum. Using inverse-variance of flux as weighting yields similar results \citep{zhu2014}. The composite spectrum $\left\langle R\left(\lambda, r_{p}\right)\right\rangle$ is the weighted average of the normalized spectra for all the galaxy-QSO pairs at a certain projected distance, which is:
\begin{equation}\label{eq:composite}
\left\langle R\left(\lambda, r_{p}\right)\right\rangle=\frac{\sum_{i} w_{i} R^{i}\left(\lambda, r_{p}\right)}{\sum_{i} w_{i}} 
\end{equation}

The above equations indicate two methods to measure the rest-frame equivalent width at each projected distance. One approach is to measure the equivalent width on a composite spectrum, we call it the stacking method for convenience. The other is named force detection, which takes the average signal of each ELG-QSO pair. Also, since the continuum fitting still biases from the intrinsic continuum of quasar spectra that leads to residuals on both stacking and force detection methods, we construct random samples to further eliminate the fitting bias. The random sample uses the real QSO spectra and shifts the spectra to randomly assigned redshift. The redshift distribution of  `foreground galaxies' is the same as our ELG sample. Also, we make sure the SNR of QSO spectra in random samples is the same as real samples at different distance ranges. 

For the stacking method, we use the average of the flux to get the composite spectrum following Equation \ref{eq:composite}. At each wavelength in the rest frame, we use 5$\sigma$ clipping to avoid outliers. To further eliminate the residual composite spectra, we use the stacking result of the random sample as the baseline of composite spectra. We measure the rest-frame equivalent width (EW) of Mg II absorption by integrating the normalized flux from 2789 $\rm{\AA}$ to 2809 $\rm{\AA}$ following Equation \ref{eq:stacking}. We check each composite spectra visually, the range of wavelength may be slightly shifted. We present the result of direct integration in the next section.

The alternative way is to measure $W_{0}^{i}$ for each galaxy-QSO pair and take the average of all the measurements as the final measurement as specified in Equation \ref{eq:force}. When we measure $W_{0}^{i}$, no prior knowledge of the absorption structure is assumed for each pair. We forcibly apply Equation \ref{eq:indi_ew} on each galaxy-QSO pair regardless of significant absorption line detection within the expected wavelength range, thus we call this method force detection. We subtract the average signal with the average equivalent width of random sample as a correction. If the continuum fitting traces the intrinsic continuum of QSO spectra, then the stacking method is equivalent to the result of force detection. Thus, these two methods serve as a cross-check. Also, the force detection method provides us with the information of absorber population.

The covering fraction $f_c$ refers to the fraction of the background light that is absorbed by the gas. It is derived by taking the ratio of absorbers with a certain equivalent width to the total number of ELG-QSO pairs. To reduce the potential biases related to continuum fitting, we utilize the covering fraction obtained from a random sample and subtract is from the covering fraction of the real sample. For details of force detection and correction, please refer to Wu24.

\section{Results}\label{basic result}
Following the method, we get the strength of Mg II absorption around ELGs spanning a distance range from 20 kpc to 10 Mpc. We also measure the Mg II absorption at different redshifts. We investigate the impact of stellar mass and SFR of ELGs on the absorption. Besides, using halo-based modeling, we fit the surface density of gas and derive the corresponding gas fraction.

\subsection{Redshift evolution of Mg II absorption}\label{abs_diffz}
\begin{figure}
\centering
\includegraphics[width=\linewidth]{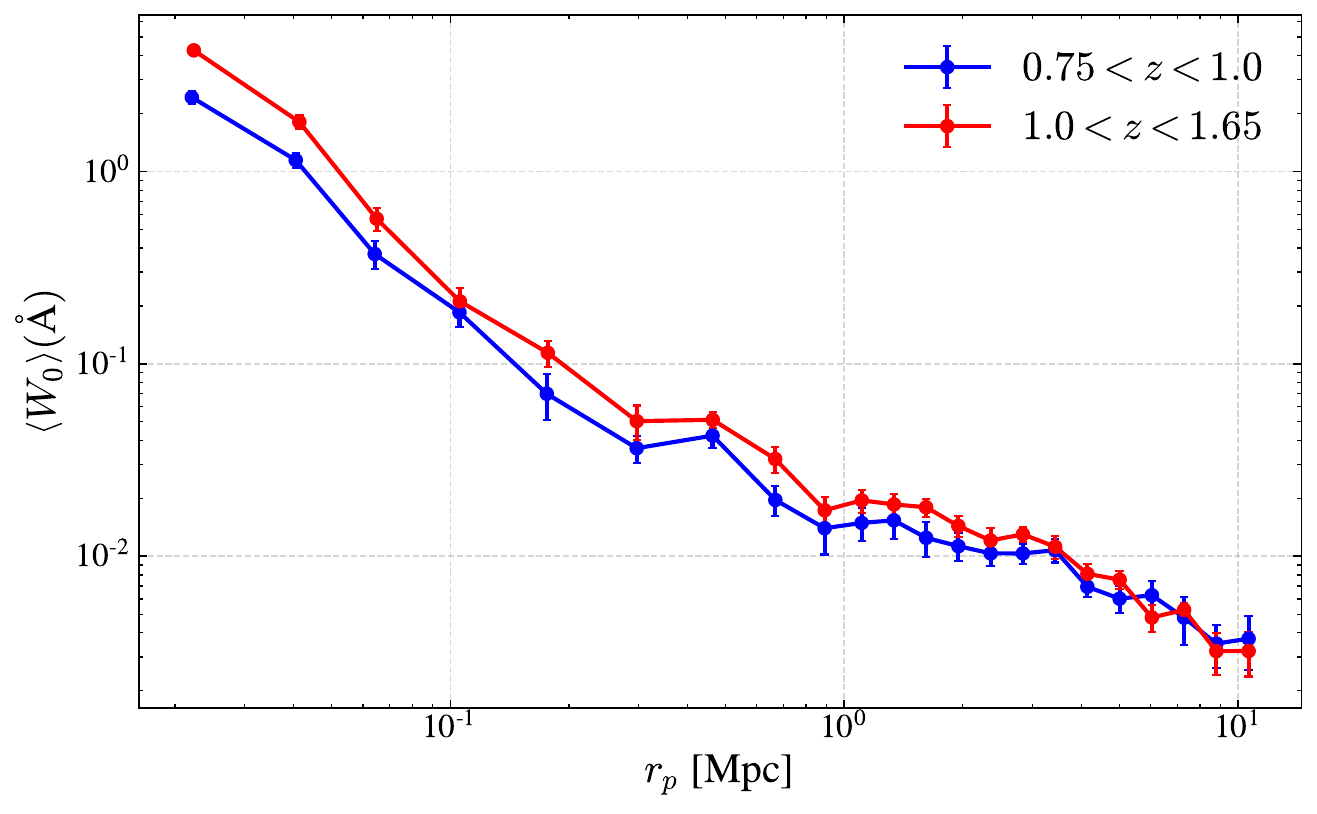}
\centering
\caption{The rest-frame total equivalent width (EW) profiles as a function of projected distance in different redshift bins. The blue line is the result for $z\in$ [0.75,1.0]. The red line is for $z\in$ [1.0,1.65]. The errors are derived by bootstrapping the measurements 100 times. MgII absorption is measured in composite spectra obtained with stacking method.}
\label{fig:zevolution}
\end{figure}

To investigate if the absorption evolves with redshift, we divide the ELGs into low and high redshift bins, which are [0.75, 1.0] and [1.0, 1.65] with median redshifts of 0.88 and 1.23. In the following discussion, we call the sample with $z<1$ as low redshift and $z>1$ as high redshift sample, respectively. The total equivalent width profiles as a function of projected distance in both redshift bins are presented in Figure \ref{fig:zevolution}. This equivalent width is derived from the stacking method. We also compile the mean EW results in Table \ref{tableofresult} along with the number of pairs in each bin in Appendix \ref{comparewith_previouswork}. The uncertainties on the equivalent widths are estimated by bootstrapping the ELG-QSO pairs 100 times. 

The total equivalent width of MgII absorbers for high redshift is consistently higher than that for low redshift sample from 20 kpc to 5 Mpc. The discrepancy becomes less apparent at $r_{p}>5$ Mpc. In the inner region, which is the scale of several hundreds of kpc, the absorption strength strongly correlates with the properties of galaxies. When the scale comes to the regime of IGM, properties of galaxy do not significantly impact the gas content. Also, the limited sample size, the treatment of normalization, and the evolution of the spectrum itself would further dilute the difference in signals.

Force detection is the average of the equivalent width of each galaxy-QSO pair, which is theoretically equivalent to the results measured from composite spectra. The rest-frame equivalent width profiles of both stacking and force detection are placed in Figure \ref{fig:methodtest} in the Appendix \ref{comparewith_previouswork}. The result of force detection is consistent with that of the stacking method, which means our procedure of fitting and stacking is valid.

\begin{figure}
\centering
\subfigure{\includegraphics[width=0.9\linewidth]{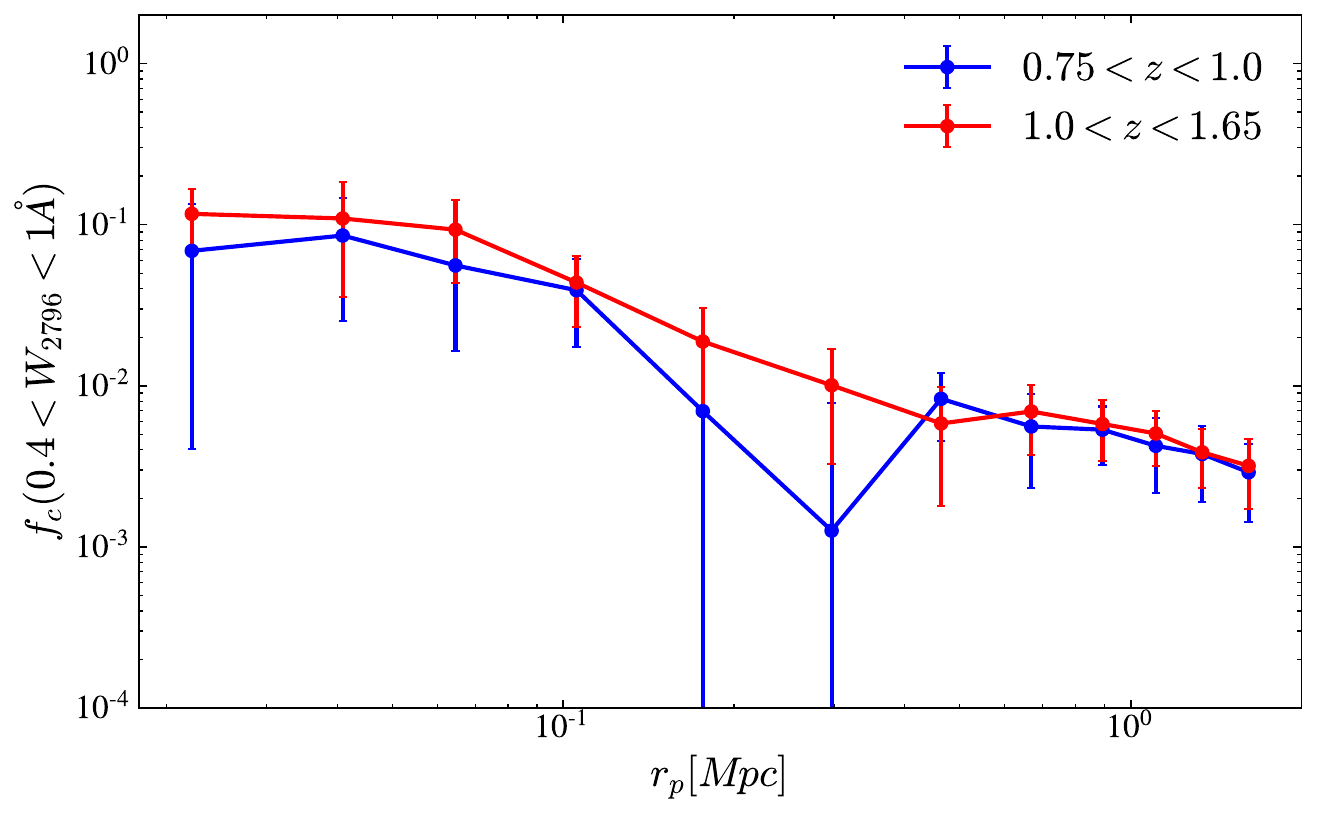}
}
\subfigure{\includegraphics[width=0.9\linewidth]{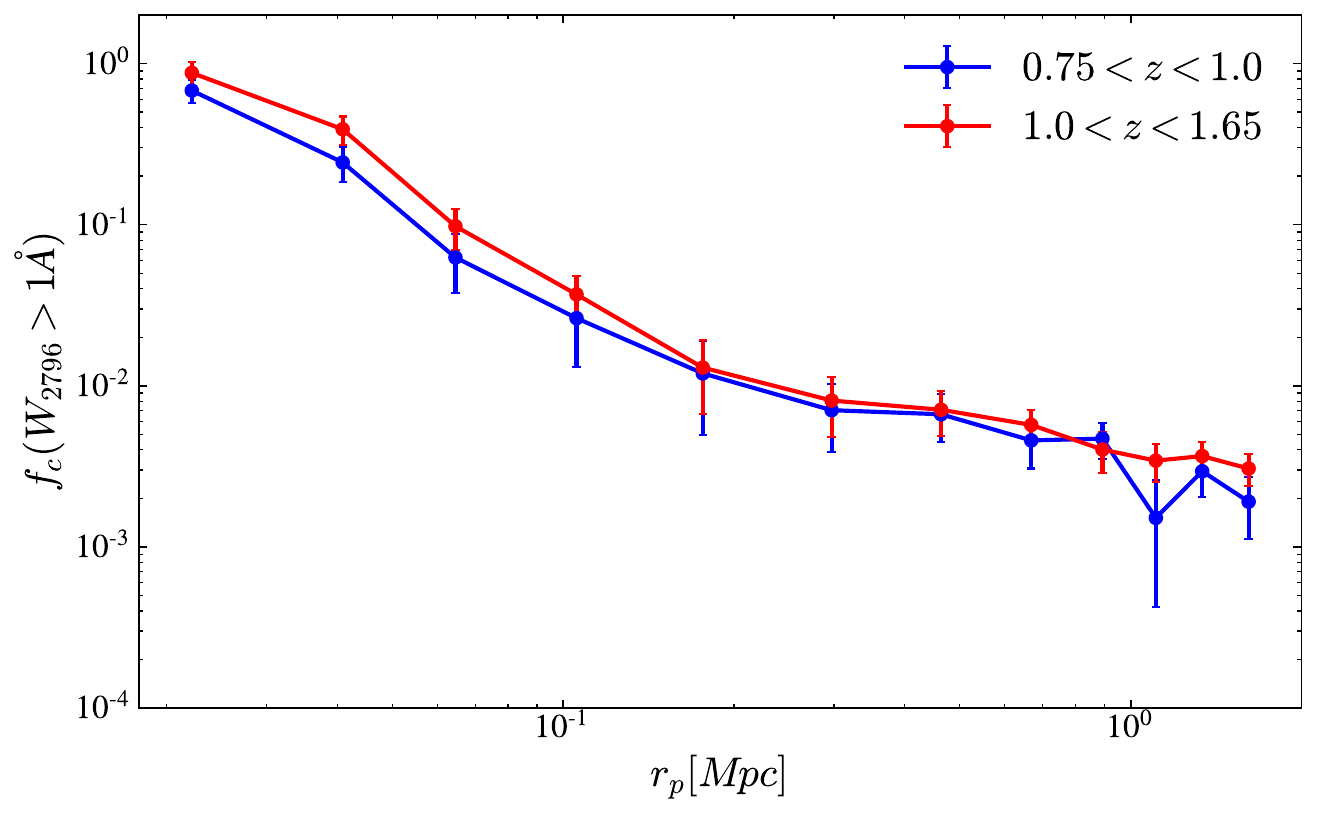}}
 
\centering
\caption{Covering fractions $f_c$ of Mg II absorbers as a function of projected distance $r_p$ for $0.4<W_{2796}<1\rm{\AA}$ and $W_{2796}>1\rm{\AA}$. Blue lines are the results for $0.75<z<1.0$ and red lines are for $1.0<z<1.65$. The errors are derived by bootstrapping the measurements 100 times.}
\label{fig:cf_tot}
\end{figure}

The covering fraction profiles are shown in Figure \ref{fig:cf_tot}. We estimate the $f_{c}$ from the random sample and use it to correct the covering fraction of the whole sample. The top panel presents the case for $0.4<W_{2796} <1\rm{\AA}$
and the bottom panel is for $W_{2796} > 1\rm{\AA}$. Here we use the physical projected distance without normalizing the distance with virial radii since the stellar mass of ELGs evolves slowly with redshift and the mean redshift virial radii of ELGs do not grow steeply.

Within about 100 kpc, $f_c$ at high redshift is systematically higher(albeit large error bars) than that at low redshift, suggesting more MgII absorbing gas systems around ELGs at higher redshift. As distance extends to hundreds of kpc, the difference in covering fractions becomes less apparent. In the outer region, the contribution from data reduction as well as the variance of spectra at different redshifts mix with the real signal. We also compare our covering fraction of $W_{2796}$ with previous works and the results are shown in Figure \ref{fig:cf_compare} in Appendix \ref{comparewith_previouswork}.

\subsection{Impact of stellar mass and SFR}\label{sfr_mass_z}
Stellar mass, SFR and gas abundance within the galactic environment exhibit evolution with redshift. Galaxies of greater stellar mass inherently contain more mass and resides in denser environments, implying a higher content of gas. We first focus on discerning the correlation between the strength of absorption and the stellar mass of galaxies.

\begin{figure*}
\centering
\includegraphics[width=\linewidth]{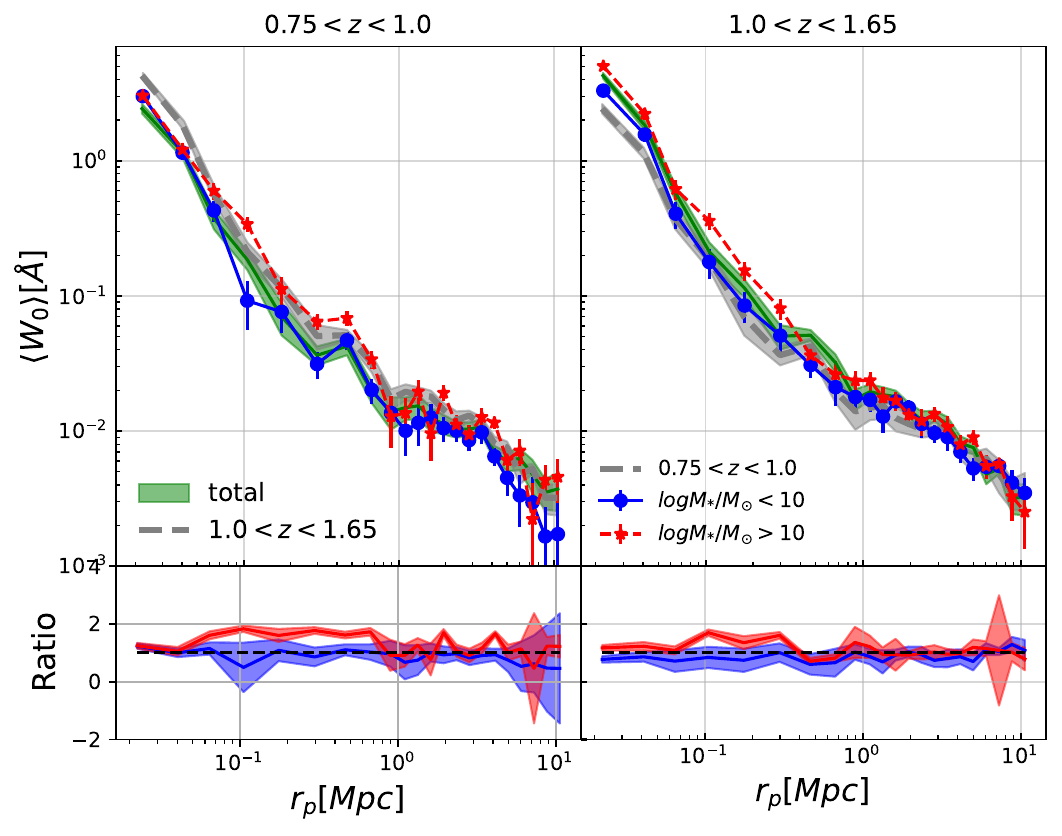}
\centering
\caption{Top panels are the profiles of total rest-frame equivalent width $W_0$ in different redshift and stellar mass $M_*$ bins. Bottom panels are the ratio of different $M_*$ and the total sample. Green lines are profiles of total sample within certain redshift ranges. Gray-shaded regions are the total results of $1.0<z<1.65$ (for the left panel) and $0.75<z<1.0$ (for right panel) which serve as references. Blue lines are results for $M_{*}<10^{10}M_{\odot}$, red lines are profiles for $M_{*}>10^{10}M_{\odot}$. The errors are derived by bootstrapping the measurements 100 times.}
\label{fig:mass}
\end{figure*}

The mean stellar mass of ELGs increases with redshift. We split ELGs into two stellar mass bins: $M_{*}<10^{10}M_{\odot}$ and $M_{*}>10^{10}M_{\odot}$ with mean stellar mass about $10^{9.7}M_{\odot}$ and $10^{10.2}M_{\odot}$ respectively. We compare the mean equivalent width of Mg II absorption within different stellar mass and redshift bins. The results are presented in Figure \ref{fig:mass}. For $z<1$, absorption around massive galaxies is stronger than that of less massive galaxies within 1 Mpc, but the enhancement disappears at $r_p > 1$ Mpc. For the higher redshift bin, the impact of stellar mass becomes less apparent, with slightly higher rest-frame equivalent width at $r_p < 1$ Mpc.

We also compare the covering fractions around ELGs with different stellar masses in Figure \ref{fig:cf_mass} in Appendix \ref{appSFRandmass}. The covering fraction of absorbers yields similar results, with more MgII systems detected around massive galaxies. Stellar mass alone does not entirely contribute to the differences in absorption strength observed at various redshift intervals. This suggests the potential influence of additional factors, such as star formation rate and the evolution of ionization states in the background.

Previous studies have revealed the co-evolution of cool gas surrounding star-forming galaxies with their star formation rates \citep{lan2020}. \citet{menard2011} find the strong Mg II traces the star formation history of galaxies up to $z \approx 2$. Additionally, \citet{anand2021} also noted an increase in the covering fraction of strong Mg II absorbers with higher SFR.

\begin{figure*}
\centering
\includegraphics[width=0.9\linewidth]{ 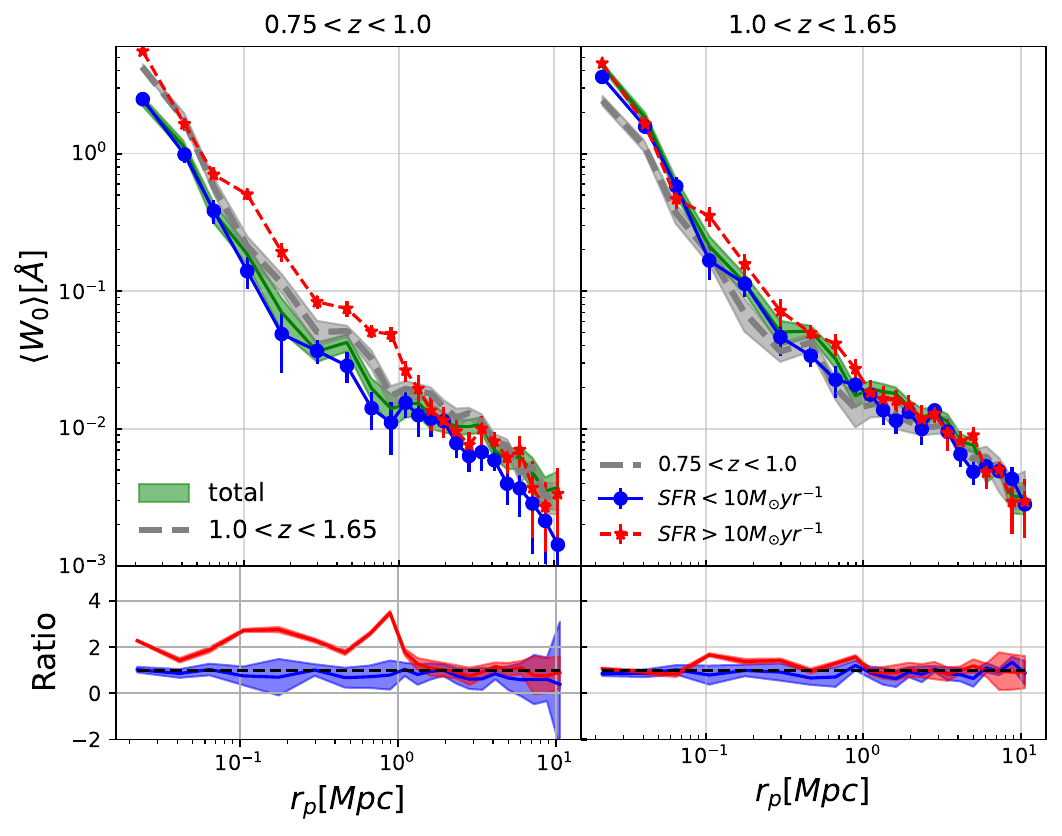}
\centering
\caption{Top panels are the profiles of rest-frame equivalent width $W_0$ at different redshift and stellar mass $M_*$. Bottom panels are the ratio of different $M_*$ and the total sample. Green lines are profiles of total sample within certain redshift ranges. Gray-shaded regions are the total results of $1.0<z<1.65$ (for the left panel) and $0.75<z<1.0$ (for the right panel) which serve as references. Blue lines are results for SFR $<$ $10$ $M_{\odot}yr^{-1}$. Red lines are for SFR $>$ $10$ $M_{\odot}yr^{-1}$. The errors are derived by bootstrapping the measurements 100 times.}
\label{fig:sfr_stack}
\end{figure*}

The mean SFR increases from low to high redshift bins. We divide galaxies into two SFR bins, which are galaxies with SFR $<$ $10$ $M_{\odot}yr^{-1}$ and SFR $>$ $10$ $M_{\odot}yr^{-1}$. We do not split the SFR into more bins due to the limitation of sample size. We present the equivalent width profiles across varying redshifts and SFR ranges in Figure \ref{fig:sfr_stack}. Remarkably, within the redshift range $z<1$, the mean absorption of cool gas demonstrates a significant enhancement with increasing SFR at projected distance $r_{p}>$1 Mpc. In fact, the equivalent width for higher SFR sample even surpasses the values observed for higher redshift bin sample.

However, this distinction between different SFR bins diminishes at higher redshift($z>1$). At higher redshift, the measured SFRs have larger uncertainties, which may dilute the intrinsic effect of SFR. We also compare the covering fractions of MgII absorbers around galaxies with lower and higher SFR in Figure \ref{fig:cf_sfr} and get similar results. Within the redshift range $z<1$, it becomes evident that the covering fraction for ELGs exhibiting high SFR is higher compared to their low SFR counterparts. However, as redshift increases, these differences attenuate, and for $z>1$, we observe only minor distinctions between the low and high SFR samples. For more details, please refer to the Appendix \ref{appSFRandmass}.

At low redshift, the effect of the properties of galaxies is remarkable. Particularly, the SFR exhibits a more prominent impact on absorption strength compared to stellar mass. However, for redshifts exceeding 1, the combined influence of both SFR and stellar mass becomes less pronounced. At high redshift, the cool gas from field and satellite galaxies contributes to the gas in the CGM galaxies \citep{Hafen2019,Faucher2023}.

\subsection{Surface density of cool gas and halo-based modeling}\label{surfacedensity_siii}

To investigate the cool gas content in the environment, we fit the cool gas fraction of IGM through halo model fitting. We first derive the surface density distribution from the absorption of Mg II. 

The surface density is estimated from the equivalent width using Equation \eqref{surfacedensity} and \eqref{columndensity}. Here $m_{\rm{ion}}$ is the mass of the certain atom, $f$ is the oscillator strength and $W_{\rm{ion}}$ is the rest-frame equivalent width of the metal absorption line at wavelength $\lambda$. The conversion assumes a linear curve of growth of equivalent width and column density, which is tenable when the absorption is unsaturated. The model is affected by the saturation effect. Saturation occurs when the column density and thermal broadening are high enough that the absorption lines become optically thick \citep{Prochaska2014}.

\begin{equation}\label{surfacedensity}
\langle \Sigma_{\rm{ion}}\rangle = m_{\rm{ion}}N_{\rm{ion}},
\end{equation}
where 
\begin{equation}\label{columndensity}
N_{\rm{ion}}= 1.13\times 10^{20}cm^{-2}\frac{W_{\rm{ion}}}{f\lambda^{2}}
\end{equation}

According to the measurement of the covering fraction, substantial signal contribution originates from strong absorbers from the IGM and CGM of galaxies rather than weak absorbers in diffused gas. The strong absorption does not follow the linear curve of growth, and the column density is higher than the value derived from Equation \ref{columndensity}. In this case, the conversion that uses the average equivalent width of Mg II only reflects the lower limit of column density.

\citet{lan2017} investigate properties of different metal lines and corresponding density. They derive the column densities of various metal absorbers from Mg II leveraging the metal absorber catalog provided by \citet{zhu2013} using spectra from the Sloan Digital Sky Survey I-III. Their analysis reveals that Si II at 1808$\rm{\AA}$ remains unsaturated even as Mg II saturates, and the redshift dependence of $W_{\rm{SiII},1808}$ and $W_{2796}$ is weak compared with other metal lines. Also, the cosmic density evolution of Si II and Mg II is close to each other. Thus we use Si II to infer the gas mass distribution in our work to reduce the effect of saturation.

We convert the equivalent width of Si II following Equation \eqref{Wsii}. Here $C=0.0034\pm 0.0005\rm{\AA}$, $\alpha=1.67\pm 0.04$ and $\beta=-0.50\pm0.15$, which are provided by \citet{lan2017}. We use the median redshift in each redshift bin to do the conversion, which is 0.88 and 1.23 respectively. At certain $W_{2796}$, we get the covering fraction and the corresponding $W_{\rm{SiII}}$. Since we are estimating the Si II in the cloud of Mg II, the covering fraction $f_c(W_{\rm{SiII}})$ is the same as $f_c(W_{2796})$. Then we derive the average column density of Si II according to Equation \eqref{averageN}. $f_c(W_{\rm{SiII}})$ and $N(W_{\rm{SiII}})$ is the covering fraction and column density at given $W_{\rm{SiII}}$. We estimate the column density of absorbers with $W_{2796}>0.4\rm{\AA}$ as stronger absorbers contribute most significantly. Weak absorbers contribute less than 1$\%$ of the total column density. Besides, the covering fractions of weak absorbers are predominantly obscured by noise. Thus the result for $W_{2796}>0\rm{\AA}$ yields a similar result. We find the slope of the column density derived from Si II exhibits a shallower profile than the slope obtained from the mean absorption of Mg II, indicating that the contribution of strong absorbers is better preserved, especially in the IGM regime. In this case, the conversion reduces the underestimation due to the saturation of strong Mg II absorbers.

\begin{equation}\label{Wsii}
W_{\rm{SiII}}= C W_{\rm{MgII},2796}^{\alpha} (1+z)^{\beta}
\end{equation}

\begin{equation}\label{averageN}
\langle N_{\rm{SiII}}\rangle = \Sigma  f_c(W_{\rm{SiII}})\times N(W_{\rm{SiII}})
\end{equation}

We assume the gas distribution roughly follows the shape of the underlying dark matter distribution and use the dark matter density profile to fit the gas distribution \citep{zhu2014}. The halo model contains a 1-halo term, which describes the gas distribution within a single halo, and a 2-halo term, which accounts for the contribution of neighboring halos. We use a projected Navarro-Frenk-White (NFW) profile \citep{Navarro1996,Navarro1997} to model the 1-halo term, which depends on halo mass, redshift, and concentration. The 2-halo term follows a projected matter-matter correlation function, which also depends on halo mass and evolves with redshift. We infer the NFW profile and matter-matter correlation function using the \texttt{python} package \texttt{COLOSSUS} \citep{COLOSSUS2018}.

The gas distribution is modeled using the following equation:
\begin{equation}\label{modelfitting}
\Sigma_{\rm{SiII}}(r_p)=\rm{f}_{1h}\Sigma_{\rm{1h}}(r_p|M_h,c,z)+\rm{f}_{2h}\Sigma_{\rm{2h}}(r_p|M_h,z),
\end{equation}
where $\rm{f}_{*h}$ represents the gas to dark matter mass ratio. The median redshift of ELGs is used for $z$. $M_h$ is the halo mass, we derive the $M_h$ using the mean value of stellar mass at each redshift bin following the stellar-to-halo mass ratio provided by \citet{shan2017}. $c$ is the concentration parameter. We calculate the concentration parameter according to halo mass and redshift \citep{shan2017}. The projected NFW profile $\Sigma_{\rm{1h}}(r_p, M_h,c)$ is used to describe the 1-halo term, while the projected halo-matter correlation function $\Sigma_{\rm{2h}}(r_p, M_h)$ represents the 2-halo term. The bias factor, which describes the excess clustering of halos over that of dark matter, is used to multiply the matter-matter correlation. We get the bias factor by using COLOSSUS for a given $M_h$ and redshift. For results at $z<1$, we fix the $M_h$ to $10^{11.55}M_{\odot}$ and central redshift to 0.88. And for results at $z>1$, we use $M_{h}=10^{11.64}M_{\odot}$ and $z=1.23$. With fixed halo mass and central redshift, we fit the surface density profiles and get the gas fraction. The fitting results are presented in Figure \ref{fig:halomodel}. In each panel, the green line is the 1-halo term fitting result while the yellow line is the 2-halo term. The red line is the combined model as described in Equation \ref{modelfitting}. For details of model fitting, please refer to Appendix \ref{apphalomodel}. 

The $\rm{f}_{1h}$ increase from $6.7\pm 1.0\times 10^{-7}$ at $z=0.88$ to $1.2\pm 0.22\times 10^{-6}$ at $z=1.23$. At higher redshift, there is more cool gas in the CGM of ELGs, which serves as the fuel for star formation. The gas fraction of the 2-halo term represents the cool gas abundance in the IGM. The abundance of cool gas traced by Si II follows:
\begin{equation}\label{omegamgii}
\Omega_{\rm{SiII}} = \Omega_{m}\times \rm{f}_{2h}
\end{equation}

\begin{figure}
\centering
\includegraphics[width=\linewidth]{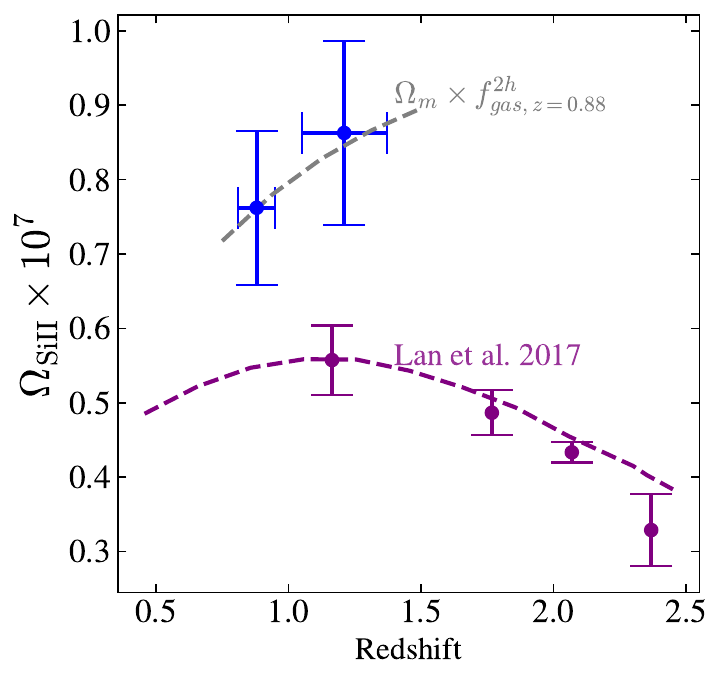}
\centering
\caption{The $\Omega_{\rm {SiII}}$ at different redshift. Errors of $\Omega_{\rm{SiII}}$ are the fitting error while errors of redshift represent the standard deviation of each redshift range. The gray dashed line shows the evolution of $\Omega_{m}$, normalized with $f^{2h}{}$Purple points and dashed line are the result of \citet{lan2017}.}
\label{fig:mgiievolution}
\end{figure}

The evolution of $\Omega_{\rm{SiII}}$ is shown in Figure \ref{fig:mgiievolution}. We put the estimation of \citet{lan2017} as a comparison. Their approach involves estimating $\Omega_{\rm{SiII}}$ based on the incidence rate of Mg II absorbers. \citet{lan2017} finds proximity in the cosmic mass density of Si II and Mg II. There is an increase in $\Omega_{\rm{SiII}}$ with redshift that mainly follows the evolution of $\Omega_{m}$. Our estimation is larger than that of \citet{lan2017} by approximately 1.5 times, with a steeper rate of increase towards higher redshift. This discrepancy is attributed to variances in the samples of absorbers since they used the absorber catalog of SDSS and the volume size. Furthermore, we are focusing on the environment around ELGs that trace high-density regions but not search for the absorbers directly within the spectra of quasars.

At higher redshift, the background metal abundance tends to be elevated due to more cool gas in the IGM at these epochs. As galaxies evolve, lower ionization states are heated to higher ionization states, leading to a decrease in such absorption features in the quasar spectra. The higher ionization states gas absorbs at even shorter wavelengths and it becomes difficult to detect them in optical at these redshifts.

\section{Summary and discussion}\label{summary}

We investigate the Mg II absorption around ELGs within the redshift intervals of $0.75<z<1.0$ and $1.0<z<1.65$. Our analysis reveals that the cool gas mean absorption in CGM and IGM evolves with redshift. Both stacking and force detection show stronger absorption in higher redshift bin, especially for $r_p<1$ Mpc. When $r_p$ extends further, the discrepancy shrinks, but the enhancement of absorption could still be seen up to $r_p\sim5$ Mpc. We also derive the covering fraction for different equivalent width thresholds, namely $0.4<W_{2796}<1\rm{\AA}$ and $W_{2796}>1\rm{\AA}$ in two redshift bins to study the evolution of the detection probability and spatial distribution of metal absorbers at different epochs. We observe that within 1 Mpc, the Mg II covering fraction is higher around ELGs at higher redshift. In the outer region, the scatter and systematics dominate the results and the differences become indistinguishable, and metal absorbers start to trace IGM rather than the CGM.

The properties of the central galaxies significantly affect the gas distribution in CGM. Specifically, the stellar mass and star formation rate (SFR) of ELGs contribute to the evolution of cool gas distribution. In the regime of IGM, the influence of hosting galaxies becomes weaker, the absorption mainly reflects the gas diffused on a large scale. 

\cite{Dutta2021} compares the dependence of Mg II absorption on galaxy properties and finds positive relations of Mg II absorption with stellar mass and SFR. They find that the covering fraction did not exhibit a strong dependence on either SFR or specific star formation rate (sSFR) within sub-samples characterized by low and high stellar mass. Besides, the influence of the galactic environment appeared to exert a more substantial impact than both stellar mass and SFR in IGM. We also find that the stellar mass and SFR only explain part of the difference as redshift increases, the environment may also impact the gas content. \cite{zhu2013} finds that the redshift path density of Mg II increases from $z=1$ to $z=2$. \cite{lan2020} compares the Mg II absorption around galaxies from $z=0.4$ to $z=1.3$ and finds strong absorbers tend to increase with redshift, which is consistent with our trend. Studies for strong absorbers at $z>2$ show that the redshift path density of strong Mg II absorbers decreases with redshift \citep{chen_sf2017,Zou2021}. The evolution of redshift path density of Mg II absorbers is consistent with the cosmic star formation history from the local universe to redshift about 6, with a turnover at about $z=2$ \citep{Madau2014}. 

To further explore the effects of the properties of galaxies, we divide our sample into different subsets based on the stellar mass. The subset with higher stellar masses exhibits a stronger absorption, indicating a higher cool gas absorption around more massive ELGs. ELGs with higher stellar masses tend to reside in denser environments, leading to a correspondingly heightened density of cool gas. Notably, the Mg II absorption is slightly higher for group than isolated system \citep{Dutta2021}. The origin of cool gas in dense environments of clusters is a complex mixture, containing processes such as gas stripping and outflows from satellite galaxies \citep{Anand2022}. In our work, we do not consider whether the ELG is a central galaxy due to the limitation of sample size. Besides, previous studies found a lower Mg II absorption in the CGM of luminous red galaxies (LRGs) than ELGs \citep{Lan2018,anand2021}. LRGs are more massive than ELGs, but most are believed to be passive galaxies with low SFR. Here we only compare the properties of ELGs and thus do not have the effect of different types of galaxies.

We compare the absorption for galaxies with different SFR. When $z<1$, the absorption around galaxies with high SFR is significantly higher than that of galaxies with low SFR. The effect of SFR becomes less conspicuous as redshift increases. The covering fractions of different SFR samples do not deviate from each other around $z>1$ galaxies. At low redshift, the SFR of galaxies has a prominent effect on the absorption around galaxies. The cool gas from the CGM of companion galaxies would also contribute to the gas absorbers we observed.

We convert the equivalent width of Mg II to Si II using the empirical relation derived in \citet{lan2017} to derive the mean column density of the cool gas, which reduces the saturation effects for strong absorbers. Notably, this conversion may suffer from intrinsic scatter due to redshift evolution. By fitting the surface density profile, we gain insights into absorption from the intergalactic medium (IGM). We calculate the cool gas surface density and employ halo model fitting to estimate the cool gas fraction. Our analysis reveals an increase in metal abundance at $z>1$, with the growth rate of Si II abundance closely tracking the overall $\Omega_{m}$, suggesting a greater presence of cool gas in the IGM at higher redshifts.  \cite{Martizzi2019} trace the evolution of gas in different phases around clusters in IllustrisTNG. They find that the gas fraction of diffuse IGM decreases from $z=2$ to $z=0$. Diffuse gas would fall into clusters and would be shock-heated to a hotter phase in the inner region of clusters. \cite{Artale2022} conduct the evolution of different ions in IllustrisTNG and find that a large amount of mass budget is in the condensed gas phase. The mass fraction of cool gas in the condensed phase increases as redshift decreases, which differs from our halo model fitting. This may relate to the AGN and stellar feedback that was added to the simulation. As galaxies evolve, more ions are heated into higher ionized states, which contributes to the observed decrease in metal abundance with redshift.

Our work builds on previous studies that have investigated the evolution of gas in the universe using absorption features in QSO spectra. We benefit from the large sample sizes of ELGs and QSOs obtained with DESI, allowing us to study the gas evolution in greater detail. We find the cool gas content increases at higher redshift, the stellar mass and SFR would partially affect the gas absorption. The diffused gas in the background also evolves with redshift. Our results provide new insights into the role of cool gas in galaxy evolution and offer important constraints on models of galaxy formation and cosmology. The strong redshift evolution of the Mg II absorption suggests a significant evolution in the amount and distribution of cool gas in the IGM over cosmic time.

Our findings also highlight the important role of the environment in shaping galaxies' properties and surroundings. Studies of $Ly\alpha$ forest absorption have provided important insights into the properties of the IGM at high redshifts. In contrast, studies of metal line absorption, such as Mg II, have probed the properties of the IGM at lower redshifts. Our work complements these studies and provides a more complete picture of the evolution of the IGM across cosmic time. Future studies could extend our work by using higher-resolution spectroscopy to study the detailed properties of the Mg II absorption and its relationship to the surrounding galaxy population. Also, the metal line forest such as CIV would trace even highly ionized gas phase, providing new insight into the gas in CGM and IGM.

\acknowledgments
We thank Matthew Pieri and Lucas Napolitano for the helpful comments in the DESI
internal review. 

We acknowledge support from the National Key R\&D Program of China (grant no.~2023YFA1605600), the National Science Foundation of China (grant no.~12073014), the science research grants from the China Manned Space Project with No.~CMS-CSST-2021-A05, and Tsinghua University Initiative Scientific Research Program (No.~20223080023). SZ acknowledges support from the National Science Foundation of China (no. 12303011).

This research is based upon work supported by the U.S. Department of Energy (DOE), Office of Science, Office of High-Energy Physics, under Contract No. DE–AC02–05CH11231, and by the National Energy Research Scientific Computing Center, a DOE Office of Science User Facility under the same contract. Additional support for DESI was provided by the U.S. National Science Foundation (NSF), Division of Astronomical Sciences under Contract No. AST-0950945 to the NSF’s National Optical-Infrared Astronomy Research Laboratory; the Science and Technology Facilities Council of the United Kingdom; the Gordon and Betty Moore Foundation; the Heising-Simons Foundation; the French Alternative Energies and Atomic Energy Commission (CEA); the National Council of Humanities, Science and Technology of Mexico (CONAHCYT); the Ministry of Science and Innovation of Spain (MICINN), and by the DESI Member Institutions: \url{https://www.desi.lbl.gov/collaborating-institutions}. Any opinions, findings, and conclusions or recommendations expressed in this material are those of the author(s) and do not necessarily reflect the views of the U. S. National Science Foundation, the U. S. Department of Energy, or any of the listed funding agencies.

The authors are honored to be permitted to conduct scientific research on Iolkam Du’ag (Kitt Peak), a mountain with particular significance to the Tohono O’odham Nation.

Data points for Table 1 and figures are available in Zenodo at https://doi.org/10.5281/zenodo.12803436

\appendix
\section{Correction for stacking and force detection}\label{comparewith_previouswork}
In this section, we further introduce our measurement and correction of stacking and force detection methods. Also, we compare our results with previous works.

\begin{figure}
\centering
\includegraphics[width=\linewidth]{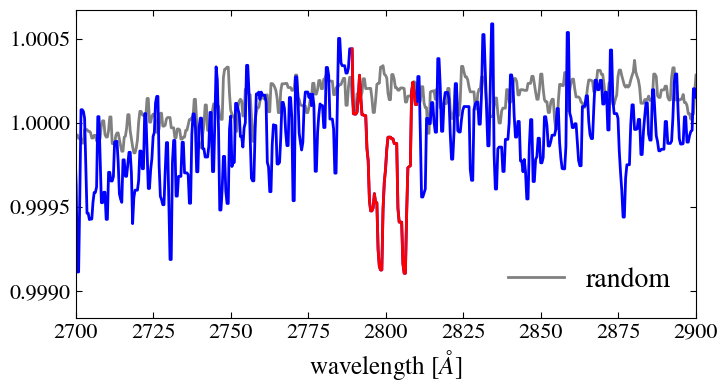}
\centering
\caption{Composite spectra for the real and random sample. The blue line is the stacking result of the real sample at $r_p=2.35Mpc$ and $z<1$. The gray line is the composite spectrum of the random sample, which is used to further eliminate the residual of the composite real sample. The red region shows the wavelength coverage of Mg II. }
\label{fig:fittingbaseline}
\end{figure}

Our continuum fitting is not able to trace the intrinsic continuum of the spectrum, after stacking spectra, there remains a systematics at the level of 0.3$\%$. As we extend to IGM, the absorption signal becomes weak to detect and the systematics would further bias our measurement of the equivalent width. In this case, we construct a composite spectrum of the random sample to eliminate the bias. The random sample uses the same set of QSO spectra and the redshift distribution of `foreground galaxies' is the same as our ELGs. By doing so, we eliminate the bias caused by normalization. However, this random sample helps to correct the systematics when there is no obvious absorption structure. Considering the low covering fraction of strong absorbers at 1 Mpc, the fitting bias around strong absorbers only accounts for a small fraction of spectra and does not have a large impact on the composite spectrum. We show one example of the composite spectra for the real and random samples in Figure \ref{fig:fittingbaseline}. The real sample is at $r_p=2.35$ Mpc and $z<1$. The random sample contains 8 million QSO spectra in total and 3.5 million spectra for $z<1$ and the shape of the random composite spectrum becomes stable as the number of spectra increases. Table \ref{tableofresult} shows the measurement of equivalent width at different redshift ranges, with the number of pairs we used in each bin.

For force detection, we also need to eliminate the fitting bias. We also generate the random sample and construct the same fitting and measurement of equivalent width on it. The SNR and redshift distribution of the foreground galaxies is the same as the real sample. The SNR of a certain QSO spectrum is the median of SNRs derived from the flux and inverse variance at each wavelength. We also calculate the SNR of each absorber, which is the SNR within the wavelength range of 2789 $\rm{\AA}$ to 2809 $\rm{\AA}$ and use the SNR to sort and derive the correction. The choice of different SNRs gives consistent results.

\begin{figure}
\centering
\includegraphics[width=\linewidth]{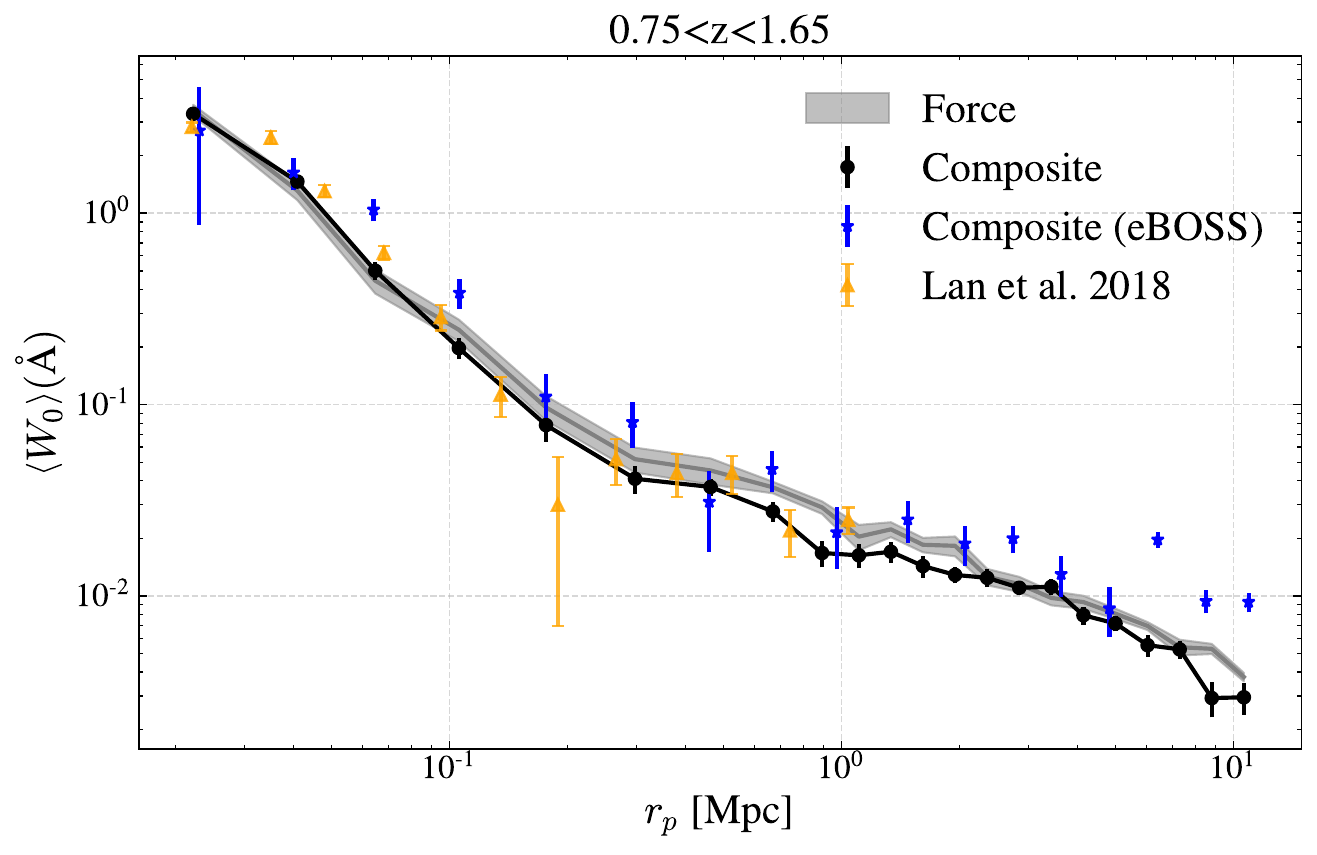}
\centering
\caption{The total rest-frame equivalent width (EW) profiles using different methods. The gray-shaded region is the result of force detection. The black dot shows the measurements of EW from the stacked QSO spectra. The errors are derived by bootstrapping the measurements 100 times. The blue dots show the result of Wu24, which uses the ELG sample of eBOSS. The yellow dots show the results of \cite{Lan2018}.}
\label{fig:methodtest}
\end{figure}

Figure \ref{fig:methodtest} illustrates the results of our stacking and force detection analyses within the redshift range z $\in$ [0.75, 1.65]. We examined the outcomes of the ELG sample from the eBOSS survey. The force detection and composite spectra show consistent results within 1 Mpc. However, at greater distances, the equivalent width (EW) derived from force detection is slightly higher than that obtained through stacking. To eliminate the normalization-related biases, we employ a random sample that is constructed from the same QSO spectra but randomly assigned the redshift of `foreground galaxies'. This guarantees that we have the same sky coverage and redshift distribution as the real sample. We also present works from \citet{Lan2018} and Wu24, both of which have independently measured the Mg II absorption around ELGs. Notably, \citet{Lan2018} utilized ELGs from the eBOSS survey and the QSO sample from SDSS DR14, while Wu24 employed ELGs from eBOSS and QSOs from SDSS DR16, with the central redshift of ELGs about 0.9. The overall trends are consistent with each other. The variation is due to systematics between SDSS and DESI spectra and different redshift coverage. However, the difference does not affect our result as we conduct a comparative analysis of absorption encompassing galaxies and QSOs only from DESI.

\begin{table}
\caption{The number of pairs and the total equivalent width of Mg II absorption in each projected distance bin at different distances. In each cell of the number and equivalent width, on the top is the result for $z<1$ and the bottom is for $z>1$.}
\centering
\begin{tabular}{ccl}\label{tableofresult}

\begin{tabular}[c]{@{}c@{}}Distance\\ {[}Mpc{]}\end{tabular} & \multicolumn{1}{l}{Number of pairs} & \begin{tabular}[c]{@{}l@{}}$W_{0}$\\\ $[\rm{\AA}]$\end{tabular} \\ \hline
\multirow{2}{*}{0.022}                                       & 45                         & 2.43$\pm$0.195                                                \\ \cline{2-3} 
                                                             & 34                         & 4.26$\pm$0.200                                                \\ \hline
\multirow{2}{*}{0.041}                                       & 75                         & 1.15$\pm$0.106                                                \\ \cline{2-3} 
                                                             & 61                         & 1.81$\pm$0.151                                                \\ \hline
\multirow{2}{*}{0.065}                                       & 150                        & 0.373$\pm$0.063                                               \\ \cline{2-3} 
                                                             & 146                        & 0.568$\pm$0.076                                               \\ \hline
\multirow{2}{*}{0.106}                                       & 495                        & 0.185$\pm$0.030                                               \\ \cline{2-3} 
                                                             & 514                        & 0.212$\pm$0.037                                               \\ \hline
\multirow{2}{*}{0.175}                                       & 1330                       & 0.070$\pm$0.019                                               \\ \cline{2-3} 
                                                             & 1447                       & 0.114$\pm$0.018                                               \\ \hline
\multirow{2}{*}{0.297}                                       & 4492                       & 0.0365$\pm$0.0059                                             \\ \cline{2-3} 
                                                             & 4674                       & 0.0504$\pm$0.0103                                             \\ \hline
\multirow{2}{*}{0.463}                                       & 11240                      & 0.0424$\pm$0.0059                                             \\ \cline{2-3} 
                                                             & 12613                      & 0.0512$\pm$0.0048                                             \\ \hline
\multirow{2}{*}{0.667}                                       & 20423                      & 0.0196$\pm$0.0034                                             \\ \cline{2-3} 
                                                             & 23433                      & 0.0320$\pm$0.0048                                             \\ \hline
\multirow{2}{*}{0.89}                                        & 32643                      & 0.0140$\pm$0.0037                                             \\ \cline{2-3} 
                                                             & 37602                      & 0.0173$\pm$0.0031                                             \\ \hline
\multirow{2}{*}{1.10}                                        & 38474                      & 0.0149$\pm$0.0029                                             \\ \cline{2-3} 
                                                             & 45606                      & 0.0195$\pm$0.0026                                             \\ \hline
\multirow{2}{*}{1.33}                                        & 56685                      & 0.0154$\pm$0.0031                                             \\ \cline{2-3} 
                                                             & 67834                      & 0.0186$\pm$0.0025                                             \\ \hline
\multirow{2}{*}{1.61}                                        & 82993                      & 0.0125$\pm$0.0025                                             \\ \cline{2-3} 
                                                             & 98995                      & 0.1798$\pm$0.0019                                             \\ \hline
\multirow{2}{*}{1.95}                                        & 120378                     & 0.0113$\pm$0.0018                                             \\ \cline{2-3} 
                                                             & 145909                     & 0.0144$\pm$0.0018                                             \\ \hline
\multirow{2}{*}{2.35}                                        & 176629                     & 0.0104$\pm$0.0014                                             \\ \cline{2-3} 
                                                             & 212203                     & 0.0121$\pm$0.0019                                             \\ \hline
\multirow{2}{*}{2.84}                                        & 257177                     & 0.0103$\pm$0.0012                                             \\ \cline{2-3} 
                                                             & 309263                     & 0.0130$\pm$0.0011                                             \\ \hline
\multirow{2}{*}{3.43}                                        & 375569                     & 0.0107$\pm$0.0015                                             \\ \cline{2-3} 
                                                             & 408271                     & 0.0112$\pm$0.0015                                             \\ \hline
\multirow{2}{*}{4.14}                                        & 545826                     & 0.0069$\pm$0.0009                                             \\ \cline{2-3} 
                                                             & 603797                     & 0.0081$\pm$0.0010                                             \\ \hline
\multirow{2}{*}{5.01}                                        & 794638                     & 0.0060$\pm$0.0009                                             \\ \cline{2-3} 
                                                             & 951263                     & 0.0076$\pm$0.0008                                             \\ \hline
\multirow{2}{*}{6.04}                                        & 1153544                    & 0.0063$\pm$0.0011                                             \\ \cline{2-3} 
                                                             & 1382903                    & 0.0048$\pm$0.0007                                             \\ \hline
\multirow{2}{*}{7.29}                                        & 1675076                    & 0.0048$\pm$0.0013                                             \\ \cline{2-3} 
                                                             & 2006117                    & 0.0052$\pm$0.0004                                             \\ \hline
\multirow{2}{*}{8.80}                                        & 2432059                    & 0.0035$\pm$0.0009                                             \\ \cline{2-3} 
                                                             & 2913366                    & 0.0032$\pm$0.0008                                             \\ \hline
\multirow{2}{*}{10.63}                                       & 3526007                    & 0.0037$\pm$0.0012                                             \\ \cline{2-3} 
                                                             & 4217720                    & 0.0032$\pm$0.0008                                             \\ \hline
\end{tabular}
\end{table}

\begin{figure}
\subfigure{\includegraphics[width=0.5\linewidth]{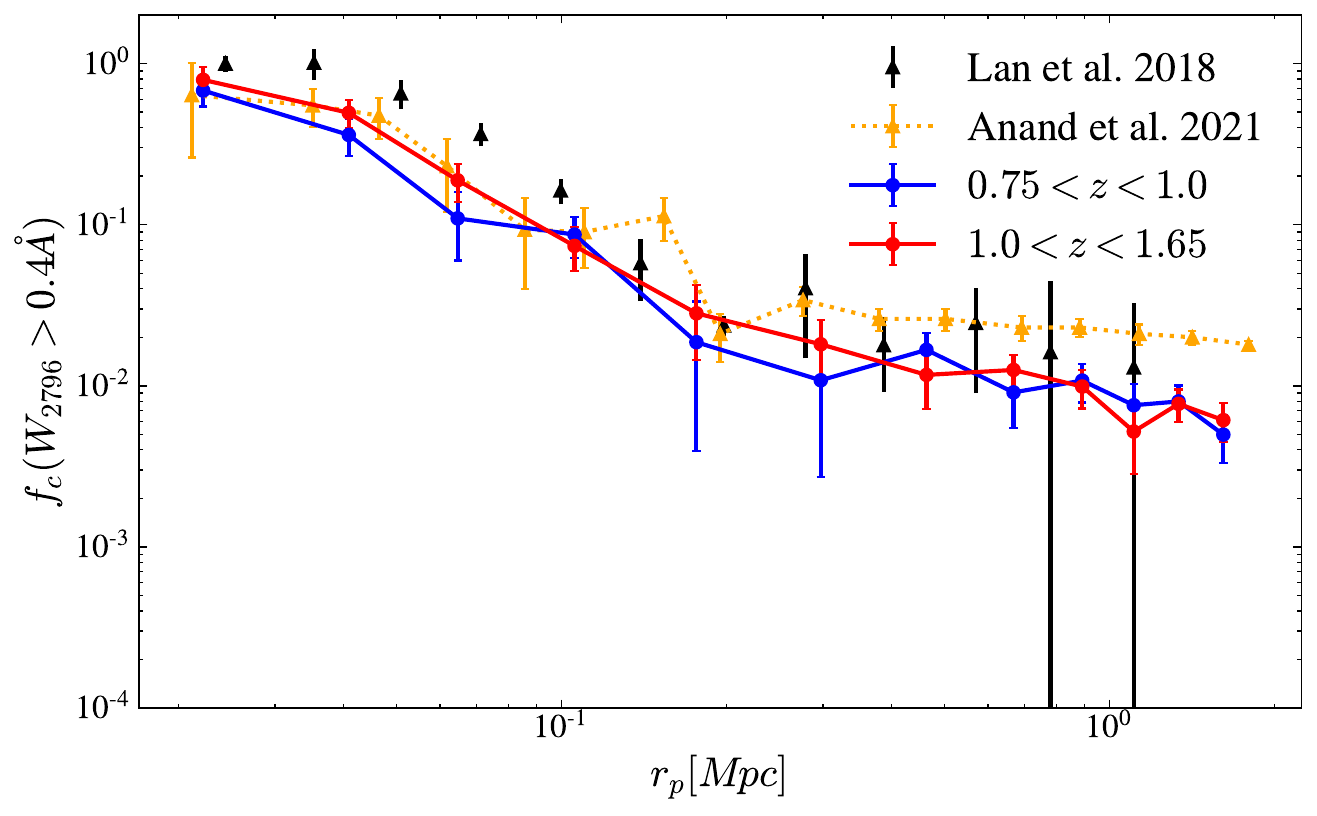}
}
\subfigure{\includegraphics[width=0.5\linewidth]{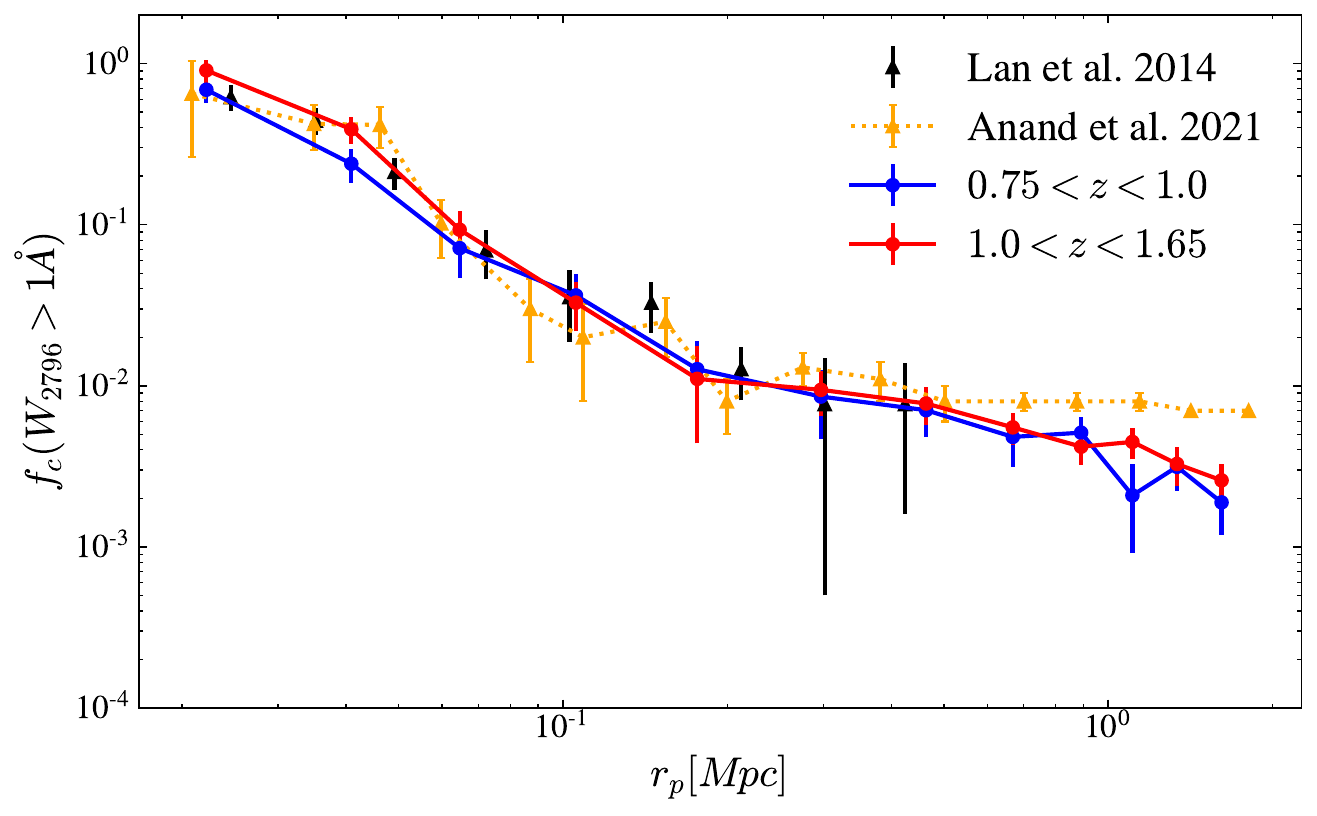}
 }
\caption{Covering fractions of Mg II absorbers as a function of projected for $W_{\lambda 2796}>0.4\rm{\AA}$ and $W_{\lambda 2796}>1\rm{\AA}$. Blue lines are the results for $0.75<z<1.0$ and red lines are for $1.0<z<1.65$. The errors are derived from 100 times bootstrapping. We compare our results with \cite{lan2014}, \cite{Lan2018}, and \cite{anand2021}, which are in black and orange respectively. }
\label{fig:cf_compare}
\end{figure}

Then we investigate the covering fraction of Mg II absorbers and compare our results with \citet{lan2014}, \citet{Lan2018}, and \citet{anand2021}  where authors have used QSOs and ELGs from SDSS. We get $W_{2796}$ from $W_0$ with the line ratio of $W_{2796}$ and $W_{2803}$ given by \citet{zhu2014}, the results are shown in Figure \ref{fig:cf_compare}. \citet{Lan2018} uses SDSS DR14 and converts the $f_c$ from the equivalent width of Mg II by assuming that the average absorption strength around galaxies can be approximated by the product of the covering fraction and mean column density of absorbers with $W_{\lambda 2796}> 0.4\rm{\AA}$. The average of $W_{2796}$ is obtained from individual absorbers. \citet{lan2014} estimate $f_c$ by counting the total number of galaxies around Mg II absorbers in a given equivalent width range divided by the total number of galaxies expected within the selected redshift range around the selected QSOs. They use the absorber of SDSS DR7 \citep{zhu2013}. \citet{anand2021} adopt SDSS DR16 \citep{lyke2020} and calculate the number of absorbers with certain absorption strengths detected in SDSS quasars. The result from \citet{anand2021} is shown in orange. We split our results into two redshift bins. Blue lines are the results for $0.75<z<1.0$ and red lines are for $1.0<z<1.65$. The errors are derived by bootstrapping the measurements 100 times. In the inner region, our results are consistent with previous works. In the outer region, \citet{anand2021} is higher than our results. The difference may relate to different approaches to getting the signal as well as different versions of data. Also, \citet{anand2021} used individual Mg II systems detected in QSO spectra and their method is sensitive to a certain equivalent width threshold. At very large distances in the IGM, the absorption is dominated by weak absorbers. Therefore, their average absorption is higher at large distances than current stacking-based results \citep{Anand2022}.

$f_c$ at higher redshift is systematically higher. At larger distances, the systematics related to data reduction would dilute the variance among different redshift intervals and we only show results within 3 Mpc, where the signal remains stronger than systematics. 

\section{Impact of stellar mass and SFR on covering fraction}\label{appSFRandmass}
\begin{figure}[h]
\includegraphics[width=\linewidth]{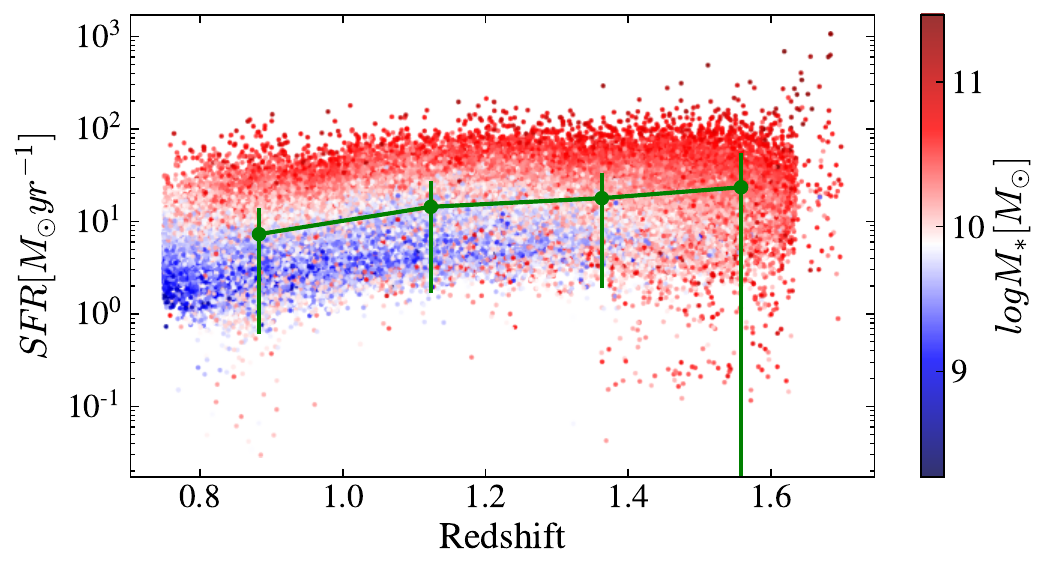}
\caption{The SFR evolution of ELGs, using color to represent different stellar mass. The green line shows the mean and scatter of SFR at different redshift ranges.}
\label{fig:z-sfr}
\end{figure}

The estimation of stellar mass is based on Random Forests which maps photometric data to Stripe 82 MGC stellar masses \citep{bundy2015}. We use EAZY \citep{Brammer2008} to estimate the SFR of ELGs using photometric data from three optical bands ($g,r,z$) and $W1$ and $W2$ bands. We also compare the stellar mass and SFR using the Code Investigating GALaxy Emission \citep[CIGALE; ][]{Boquien2019}, the estimation is provided by M. Siudek et al. in prep. The results of different methods follow similar trends, thus the choice of methods does not affect our result. In Figure \ref{fig:z-sfr} we show the distribution of SFR at different redshifts, color-coded by stellar mass of galaxies. The median value of SFR rises from approximately 7 to 20$M_{\odot}yr^{-1}$ as redshift increases from 0.8 to 1.6. And the average stellar mass extends from $10^{9.8}M_{\odot}$ to $10^{10.3}M_{\odot}$. 

The stellar mass of ELG is estimated based on Random Forests \citep{breiman2001} that maps photometric data from three optical bands (
$g,r,z$) as well as WISE $W1$ and $W2$ to Stripe 82 MGC stellar masses \citep{bundy2015}. For details of the methodology, please refer to \citet{Zhourp2023}. The average stellar mass of our ELG sample extends from $10^{9.8}M_{\odot}$ to $10^{10.3}M_{\odot}$ as the redshift changes from 0.8 to 1.6. 

We estimate the SFR of ELGs using EAZY \citep{Brammer2008}, incorporating photometric data from three optical bands (
$g,r,z$) and $W1$ and $W2$ bands for each galaxy.

\begin{figure}
\centering
\includegraphics[width=\linewidth]{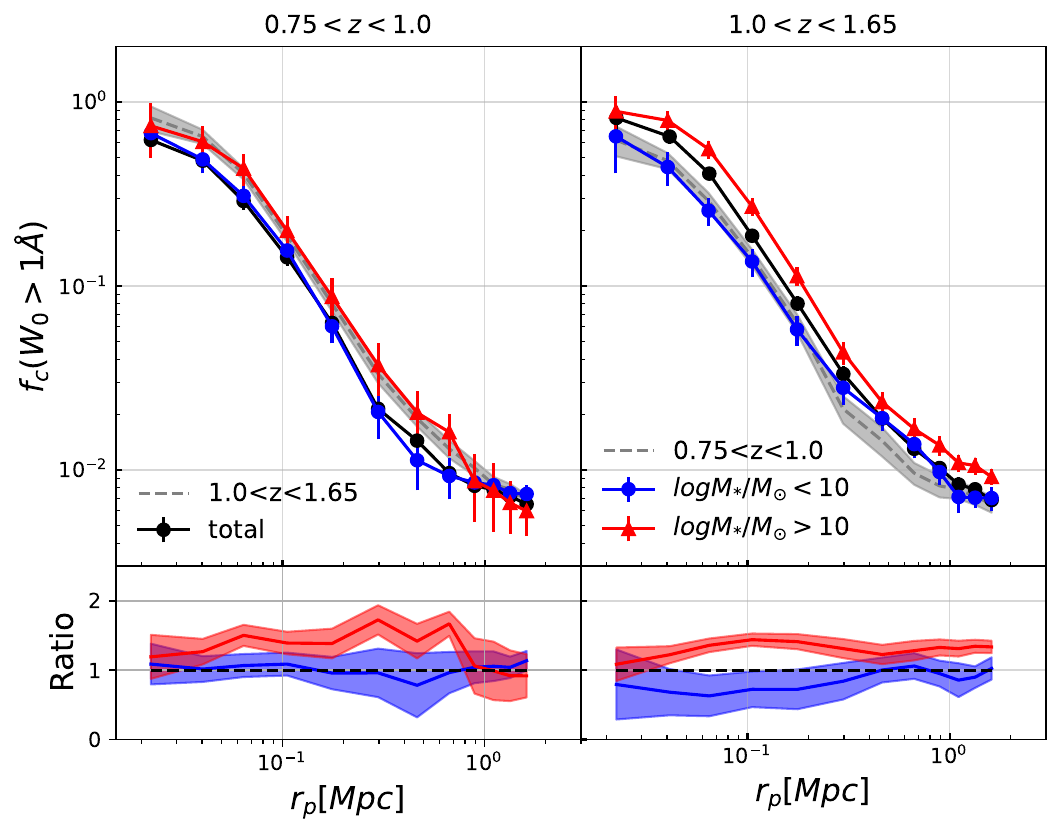}

\centering
\caption{Covering fraction $f_c$ for absorbers with total equivalent width $W_{0}>1\rm{\AA}$. Top panels are $f_c(W_0 >1\rm{\AA})$ profiles at different stellar mass and redshift ranges. Bottom panels are the ratio of different $M_{*}$ and the total sample. Blue lines are for $M_{*}<10^{10}M_{\odot}$, red lines in the left panel are profiles for $M_{*}>10^{10}M_{\odot}$. Black lines are profiles of total samples within certain redshift ranges. The errors are derived by bootstrapping the measurements 100 times. }
\label{fig:cf_mass}
\end{figure}

We explore the influence of stellar mass on the covering fraction $f_c$. Figure \ref{fig:cf_mass} displays the covering fraction across different redshift bins, with the dataset divided into different stellar mass bins in each redshift range. To make a comparison among sub-samples, we normalize $f_c$ using the relative number of galaxies within each stellar mass bin. For the low redshift sample, there is a systematic evolution around galaxies with $M_{*}>10^{10}M_{\odot}$ compared to galaxies with $M_{*}<10^{10}M_{\odot}$, with a more pronounced enhancement for strong absorbers. As redshift increases, the distinction becomes less significant. This disparity converges as the analysis extends to 1 Mpc.

\begin{figure}
\centering
\includegraphics[width=\linewidth]{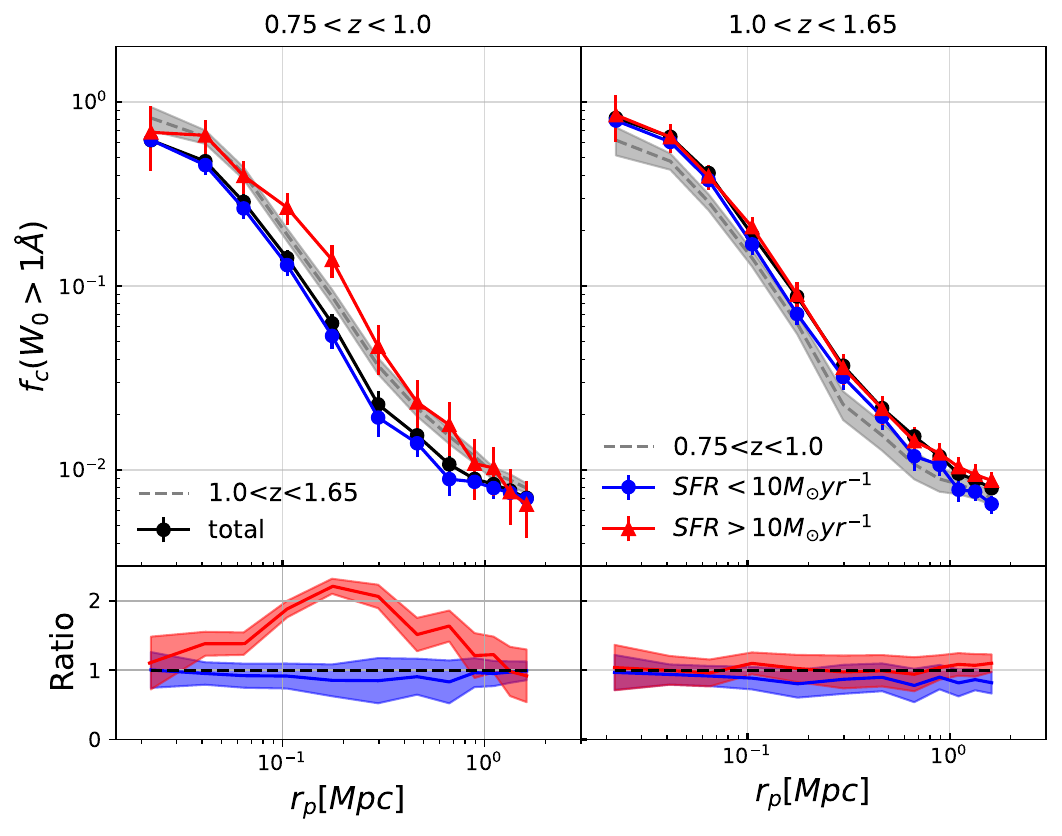}
\centering
\caption{Covering fraction $f_c$ for absorbers with equivalent width $W_{0}>1\rm{\AA}$. Top panels are $f_c(W_0 >1\rm{\AA})$ profiles at different stellar mass and redshift ranges. Bottom panels are the ratio of different SFR and the total sample. Blue lines are results for SFR $<$ $10$ $M_{\odot}yr^{-1}$, red lines are profiles for SFR $>$ $10$ $M_{\odot}yr^{-1}$ at $z>1$. Black lines are profiles of total sample within certain redshift ranges. The errors are derived by bootstrapping the measurements 100 times. }
\label{fig:cf_sfr}
\end{figure}

Then we investigate the evolution of covering fractions with different SFR. Figure \ref{fig:cf_sfr} shows the covering fraction with $W_0>1\rm{\AA}$ for galaxies within different SFR bins. Here we use the total equivalent width $W_0$ to reduce the scatter brought in as we need to use line ratio to calculate $W_{2796}$. 

We compare the ratio of different SFR ranges. For the low redshift sample, the impact on SFR is significant, galaxies with high SFR exhibit a higher covering fraction of Mg II absorbers. For the high redshift sample, the differences disappear and the influence of SFR on cool gas covering fraction seems not significant. At high redshift, the estimation of SFR contains higher uncertainty, this would further dilute the disparity.

\section{Halo model fitting}\label{apphalomodel}
In this section, we show the results of the halo model fitting. We assume the distribution of gas mass mainly follows the distribution of dark matter around galaxies, with gas fractions left as free parameters. The model contains a 1-halo term, which describes the gas distribution within a single halo, and a 2-halo term, which is for the contribution from neighboring halos. We use Bayesian inference to constrain the gas fraction on 1-halo term and 2-halo term. Here we estimate the parameters using PyMultinest, which is a Python module of MultiNest sampling that provides a fast estimation of parameters \citep[]{multinest}. For priors, we assume they follow a flat distribution that is large enough to cover each quantity. The green line is the 1-halo term fitting result while the yellow line is the 2-halo term. The red line is the combined model fitting with both 1-halo and 2-halo term. The errors are conveyed from the standard deviation of the posterior distributions of gas fractions. 
\begin{figure}
\subfigure{\includegraphics[width=0.5\linewidth]{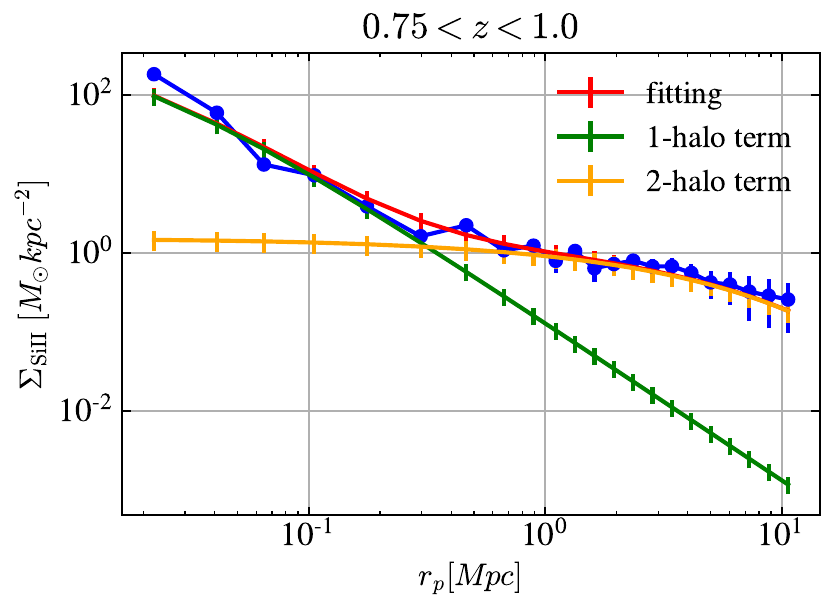}
}
 \subfigure{\includegraphics[width=0.5\linewidth]{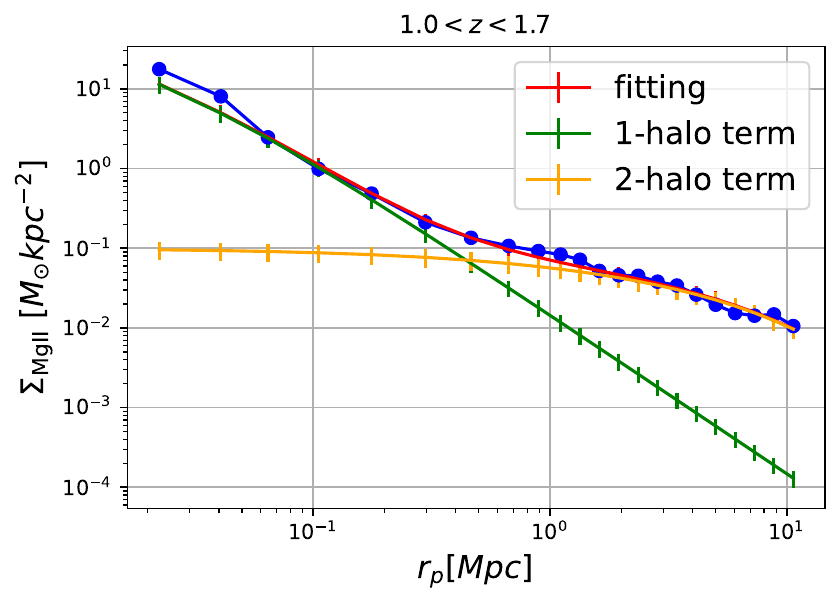}
 }
\caption{The halo model fitting results. From left to right panels, the results are $z\in$ [0.75,1.0] and $z\in$ [1.0,1.65] respectively. In each panel, the blue line is the surface density at a certain redshift. The red line is the model fitting. The green line is the 1-halo term fitting result while the yellow line is the 2-halo term.}
\label{fig:halomodel}
\end{figure}

\bibliographystyle{aasjournal}
\bibliography{references}

\end{document}